\documentclass[12pt,journal,onecolumn]{IEEEtran}
\usepackage{epsfig,setspace,amsmath,epsf,amssymb,bm,theorem,cite, graphicx, epstopdf, algorithm, algpseudocode,float,color}
\usepackage{multirow}
\usepackage[table,xcdraw]{xcolor}

\newtheorem{theorem}{Theorem}
\newtheorem{corollary}{Corollary}

\newtheorem{lemma}{Lemma}

\newenvironment{Proof}[1]{\medskip\par\noindent{\bf Proof:\,}\,#1}{{\mbox{\,$\blacksquare$}\par}}

\allowdisplaybreaks

\begin{document}	
\title{Cache-Aided Private Information Retrieval with Partially Known Uncoded Prefetching: Fundamental Limits\thanks{This work was supported by NSF Grants CNS 13-14733, CCF 14-22111, CNS 15-26608 and CCF 17-13977.}}
	
\author{Yi-Peng Wei \qquad Karim Banawan \qquad Sennur Ulukus\\
	\normalsize Department of Electrical and Computer Engineering\\
	\normalsize University of Maryland, College Park, MD 20742 \\
	\normalsize {\it ypwei@umd.edu  \qquad \it kbanawan@umd.edu} \qquad {\it ulukus@umd.edu}}
	
\maketitle
	
\vspace*{-1.2cm}

\begin{abstract}
We consider the problem of private information retrieval (PIR) from $N$ non-colluding and replicated databases, when the user is equipped with a cache that holds an uncoded fraction $r$ of the symbols from each of the $K$ stored messages in the databases. This model operates in a two-phase scheme, namely, the prefetching phase where the user acquires side information and the retrieval phase where the user privately downloads the desired message. In the prefetching phase, the user receives $\frac{r}{N}$ uncoded fraction of each message from the $n$th database. This side information is known only to the $n$th database and unknown to the remaining databases, i.e., the user possesses \emph{partially known} side information. We investigate the optimal normalized download cost $D^*(r)$ in the retrieval phase as a function of $K$, $N$, $r$. We develop lower and upper bounds for the optimal download cost. The bounds match in general for the cases of very low caching ratio ($r \leq \frac{1}{N^{K-1}}$) and very high caching ratio ($r \geq \frac{K-2}{N^2-3N+KN}$). We fully characterize the optimal download cost caching ratio tradeoff for $K=3$. For general $K$, $N$, and $r$, we show that the largest gap between the achievability and the converse bounds is $\frac{5}{32}$.
\end{abstract}
	
\section{Introduction}

In today's communication networks, the end-users are equipped with large memories, and the data transmitted in the network has shifted from real-time generated data like voice to pre-generated content like movies. These two factors together have enabled caching techniques, which store data in user cache a priori in order to reduce the peak-hour network traffic load. In the meanwhile, privacy has become an important consideration for users, who wish to download data from publicly accessable databases as privately and as efficiently as possible. This is studied under the subject of private information retrieval (PIR). In this paper, we combine the caching and PIR approaches, and consider the problem of PIR for a cache-enabled end-user.

The PIR problem considers the privacy of the requested message by a user from distributed databases. In the classical setting of PIR \cite{ChorPIR}, there are $N$ non-communicating databases, each storing the same set of $K$ messages. The user wishes to download one of these $K$ messages without letting the databases know the identity of the desired message. A feasible scheme is to download all the $K$ messages from a database. However, this results in excessive download cost since it results in a download that is $K$ times the size of the desired message. The goal of the PIR problem is to construct an efficient retrieval scheme such that no database knows which message is retrieved. The PIR problem has originated in computer science \cite{ChorPIR, PIRsurvey2004, cachin1999computationally, ostrovsky2007survey, yekhanin2010private} and has drawn significant attention in information theory \cite{RamchandranPIR, unsynchonizedPIR, YamamotoPIR, VardyConf2015, RazanPIR, JafarConf2016} in recent years.

Recently, Sun and Jafar \cite{JafarPIR} have characterized the optimal normalized download cost for the classical PIR problem to be $\frac{D}{L}=\left( 1+\frac{1}{N}+\dots+\frac{1}{N^{K-1}}\right)$, where $L$ is the message size and $D$ is the total number of downloaded bits from the $N$ databases. Since the work of Sun-Jafar \cite{JafarPIR}, many interesting variants of the classical PIR problem have been investigated, such as, PIR from colluding databases, robust PIR, symmetric PIR, PIR from MDS-coded databases, PIR for arbitrary message lengths, multi-round PIR, multi-message PIR, PIR from Byzantine databases, secure symmetric PIR with adversaries, cache-aided PIR, PIR with private side information (PSI), PIR for functions, storage constrained PIR, and their several combinations 
\cite{JafarColluding,symmetricPIR,KarimCoded,arbmsgPIR,codedsymmetric,MultiroundPIR,codedcolluded,codedcolludedJafar,arbitraryCollusion,MPIRjournal,codedcolludingZhang,MPIRcodedcolludingZhang,BPIRjournal,symmetricByzantine,tandon2017capacity,wang2017linear,kadhe2017private,wei2017fundamental,chen2017capacity,wei2017capacity,sun2017_computation,mirmohseni2017private,abdul2017private}. 

% JafarColluding, May 2, 2016
% symmetricPIR, Jun. 28, 2016
% KarimCoded, Sep. 26, 2016  
% arbmsgPIR, Oct. 10, 2016
% codedsymmetric, Oct. 14, 2016
% MultiroundPIR, Nov. 7, 2016
% codedcolluded, Nov. 7, 2016
% codedcolludedJafar, Jan. 26, 2017
% arbitraryCollusion, Jan. 26, 2017
% Mpirjournal, Feb. 6, 2017 
% codedcolludingZhang, Apr. 22, 2017
% MpircodedcolludingZhang, May. 9, 2017
% Bpirjournal, Jun. 5, 2017
% symmetricByzantine, Jul. 7, 2017
% tandon2017capacity, Jun. 21, 2017
% wang2017linear, Aug. 17, 2017
% kadhe2017private, Sep. 1, 2017
% wei2017fundamental, Sep. 4, 2017
% chen2017capacity, Spe. 10, 2017
% wei2017capacity, Oct. 2, 2017
% sun2017_computation, Oct. 30, 2017
% mirmohseni2017private, Nov. 13, 2017
% abdul2017private, Nov. 14, 2017

Also recently, Maddah-Ali and Niesen \cite{maddah2014fundamental} have proposed a theoretical framework to study the tradeoff between the cache memory size of users and the network traffic load for a two-phase scheme. In the prefetching phase, when the network traffic is low, the server allocates data to the user's cache memory. In the retrieval phase, when the network traffic is high, the server delivers the messages according to the users' requests. By jointly designing the prefetching of the cache content during the low traffic period and the delivery of the requested content during the high traffic period, the coded-caching technique proposed in \cite{maddah2014fundamental} achieves a global caching gain significantly reducing the peak-time traffic load. The concept of coded-caching has been applied to many different scenarios, such as, decentralized networks, device to device networks, random demands, online settings, general cache networks, security constraints, finite file size constraints, broadcast channels and their several combinations \cite{maddah2015decentralized,pedarsani2016online,sengupta2015fundamental,ji2016fundamental,ghasemi2017improved,ji2017order,timo2015joint,shanmugam2016finite,zhang2017fundamental,gregori2016wireless,amiri2017fundamental,tian2016caching,wan2016optimality,yu2016exact,zewail2016fundamental,naderializadeh2017optimality,bidokhti2017benefits,tang2017low,xiang2017cache}.

% maddah2015decentralized, Jan. 24, 2013
% pedarsani2016online, Nov. 14, 2013
% sengupta2015fundamental, Dec. 13, 2013
% ji2016fundamental, May. 21, 2014
% ghasemi2017improved, Jan. 24, 2015
% ji2017order, Feb. 10, 2015
% timo2015joint, Aug. 2015
% shanmugam2016finite, Aug. 21, 2015
% zhang2017fundamental, Nov. 12, 2015
% gregori2016wireless, Mar. 14, 2016
% amiri2017fundamental, Apr. 13, 2016
% tian2016caching, Apr. 28, 2016
% wan2016optimality, Sep. 2016
% yu2016exact, Sep. 25, 2016
% zewail2016fundamental, Oct. 2016
% naderializadeh2017optimality, Jan. 20, 2017
% bidokhti2017benefits, Feb. 26, 2017
% tang2017low, May. 31, 2017
% xiang2017cache, Nov. 7, 2017

The caching technique not only reduces the traffic load but can also help the user to privately retrieve the desired file more efficiently by providing additional side information. The interplay between side information and the PIR problem has been studied recently in \cite{tandon2017capacity, kadhe2017private, wei2017fundamental, chen2017capacity, wei2017capacity}. We first recall that the achievability scheme proposed in \cite{JafarPIR} is based on three principles: database symmetry, message symmetry, and side information utilization. The side information in \cite{JafarPIR} comes from the undesired bits downloaded from the other $(N-1)$ databases. Side information plays an important role in the PIR problem; for instance, when $N=1$ (single database), i.e., when no side information is available, the normalized download cost is $K$, which is the largest possible. Caching can improve PIR download cost by providing useful side information.

Reference \cite{tandon2017capacity} is the first to study the cache-aided PIR problem. In \cite{tandon2017capacity}, the user has a memory of size $KLr$ bits and can store an arbitrary function of the $K$ messages, where $0\leq r\leq 1$ is the caching ratio. \cite{tandon2017capacity} considers the case when the cached content is fully known to all $N$ databases, and determines the optimal normalized download cost to be $(1-r) \left(1+\frac{1}{N}+\dots+\frac{1}{N^{K-1}}\right)$. Although the result is pessimistic since it implies that the user cannot utilize the cached content to further reduce the download cost, \cite{tandon2017capacity} reveals two new dimensions for the cache-aided PIR problem. The first one is the databases' awareness of the side information at its initial acquisition. Different from \cite{tandon2017capacity}, \cite{kadhe2017private, wei2017fundamental, chen2017capacity} study the case when the databases are unaware of the cached side information, and \cite{wei2017capacity} studies the case when the databases are partially aware of the cached side information. The second one is the structure of the side information. Instead of storing an arbitrary function of the $K$ messages, \cite{kadhe2017private, chen2017capacity, wei2017capacity} consider caching $M$ full messages out of total $K$ messages, and  \cite{wei2017fundamental} considers storing an $r$ fraction of each message in uncoded form.

This paper is closely related to \cite{wei2017fundamental}. In \cite{wei2017fundamental}, the databases are assumed to be completely unaware of the side information. However, this may be practically challenging to implement. Here, we consider a more natural model which uses the same set of databases for both prefetching and retrieval phases. Therefore, different from \cite{wei2017fundamental}, here each database gains partial knowledge about the side information, that is the part it provides during the prefetching phase. Our aim is to determine if there is a rate loss due to this partial knowledge with respect to the fully unknown case in \cite{wei2017fundamental}, and characterize this rate loss as a function of $K$, $N$ and $r$.

In this work, we consider PIR with \emph{partially known uncoded prefetching}. We consider the PIR problem with a two-phase scheme, namely, prefetching phase and retrieval phase. In the prefetching phase, the user caches an uncoded $\frac{r}{N}$ fraction of each message from the $n$th database. The $n$th database is aware of these $\frac{KLr}{N}$ bit side information, while it has no knowledge about the cached bits from the other $(N-1)$ databases. We aim at characterizing the optimal tradeoff between the normalized download cost $\frac{D(r)}{L}$ and the caching ratio $r$. For the outer bound, we explicitly determine the achievable download rates for specific $K+1$ caching ratios. Download rates for any other caching ratio can be achieved by memory-sharing between the nearest explicit points. Hence, the outer bound is a piece-wise linear curve which consists of $K$ line segments. For the inner bound, we extend the techniques of \cite{JafarPIR,wei2017fundamental} to obtain a piece-wise linear curve which also consists of $K$ line segments. We show that the inner and the outer bounds match exactly  at three line segments for any $K$.  Consequently, we characterize the optimal tradeoff for the very low ($r \leq \frac{1}{N^{K-1}}$) and the very high ($r \geq \frac{K-2}{N^2-3N+KN}$) caching ratios. As a direct corollary, we fully characterize the optimal download cost caching ratio tradeoff for $K=3$ messages. For general $K$, $N$ and $r$, we show that the worst-case gap between the inner and the outer bounds is $\frac{5}{32}$.

\section{System Model}
We consider a PIR problem with $N$ non-communicating databases. Each database stores an identical copy of $K$ statistically independent messages, $W_1, \dots, W_K$. Each message is $L$ bits long,
\begin{align}
H(W_1)=\dots=H(W_K)=L, \qquad H(W_1, \dots, W_K)=H(W_1)+\dots+H(W_K).
\end{align}
The user (retriever) has a local cache memory which can store up to $KLr$ bits, where $0\leq r\leq1$, and $r$ is called the \textit{caching ratio}. There are two phases in this system: the \textit{prefetching phase} and the  \textit{retrieval phase}.

In the prefetching phase, for each message $W_k$, the user randomly and independently chooses $Lr$ bits out of the $L$ bits to cache. The user caches the $Lr$ bits of each message by prefetching the same amount of bits from each database, i.e., the user prefetches $\frac{KLr}{N}$ bits from each database. $\forall n\in [N]$, where $[N]= \{1,2, \dots, N\}$, we denote the indices of the cached bits from the $n$th database by $\mathbb{H}_n$ and the cached bits from the $n$th database by the random variable $Z_n$. Therefore, the overall cached content $Z$ is equal to $(Z_1, \dots, Z_N)$, and $H(Z)=\sum_{n=1}^N H(Z_n)=KLr$. We further denote the indices of the cached bits by $\mathbb{H}$. Therefore, we have $\mathbb{H}= \bigcup_{n=1}^N \mathbb{H}_n$, where $\mathbb{H}_{n_1} \cap \mathbb{H}_{n_2} = \emptyset$, if $n_1 \neq n_2$. Since the user caches a subset of the bits from each message, this is called \textit{uncoded prefetching}. Here, we consider the case where database $n$ knows $\mathbb{H}_n$, but it does not know $\mathbb{H} \setminus \mathbb{H}_n$. We refer to $Z$ as \textit{partially known prefetching}.

In the retrieval phase, the user privately generates an index $\theta \in [K]$, and wishes to retrieve message $W_\theta$ such that it is impossible for any individual database to identify $\theta$. Note that during the prefetching phase, the desired message is unknown a priori. Therefore, the cached bit indices $\mathbb{H}$ are independent of the desired message index $\theta$. Note further that the cached bit indices $\mathbb{H}$ are independent of the message contents. Therefore, for random variables $\theta$, $\mathbb{H}$, and $W_1,\dots,W_K$, we have
\begin{align} \label{independency}
H\left(\theta, \mathbb{H}, W_1,\dots,W_K  \right)= H\left( \theta \right) + H\left( \mathbb{H} \right) + H(W_1)+\dots+H(W_K).
\end{align}

The user sends $N$ queries $Q_1^{[\theta]}, \dots, Q_N^{[\theta]}$ to the $N$ databases, where $Q_n^{[\theta]}$ is the query sent to the $n$th database for message $W_\theta$. The queries are generated according to $\mathbb{H}$, which are independent of the realizations of the $K$ messages. Therefore,
\begin{align} \label{query_indep}
I(W_1, \dots, W_K;  Q_1^{[\theta]}, \dots,  Q_N^{[\theta]}  ) =0.
\end{align}
To ensure that individual databases do not know which message is retrieved, we need to satisfy the following privacy constraint, $\forall n \in [N]$, $\forall \theta \in [K]$,
\begin{align} \label{privacy_constraint}
 (Q_n^{[1]}, A_n^{[1]}, W_1, \dots, W_K, \mathbb{H}_n)   \sim (Q_n^{[\theta]}, A_n^{[\theta]}, W_1, \dots, W_K, \mathbb{H}_n) ,
\end{align}
where $A \sim B$ means that $A$ and $B$ are identically distributed.

After receiving the query $Q_n^{[\theta]}$, the $n$th database replies with an answering string $A_n^{[\theta]}$, which is a function of  $Q_n^{[\theta]}$ and all the $K$ messages. Therefore, $\forall \theta \in [K], \forall n \in [N]$,
\begin{align} \label{answer_constraint}
H(A_n^{[\theta]}|Q_n^{[\theta]}, W_1, \dots, W_K)=0.
\end{align}
After receiving the answering strings $A_1^{[\theta]}, \dots, A_N^{[\theta]}$ from all the $N$ databases, the user needs to decode the desired message $W_\theta$ reliably. By using Fano's inequality, we have the following reliability constraint
\begin{align} \label{reliability_constraint}
H\left(W_\theta|Z, \mathbb{H}, Q_1^{[\theta]}, \dots, Q_N^{[\theta]}, A_1^{[\theta]}, \dots, A_N^{[\theta]} \right) = o(L),
\end{align}
where $o(L)$ denotes a function such that $\frac{o(L)}{L} \rightarrow 0$ as $L \rightarrow \infty$.

For a fixed $N$, $K$, and caching ratio $r$, a pair $\left(D(r),L\right)$ is achievable if there exists a PIR scheme for message of size $L$ bits long with partially known uncoded prefetching satisfying the privacy constraint \eqref{privacy_constraint} and the reliability constraint \eqref{reliability_constraint}, where $D(r)$ represents the expected number of downloaded bits (over all the queries) from the $N$ databases via the answering strings $A_{1:N}^{[\theta]}$, where $A_{1:N}^{[\theta]}=(A_1^{[\theta]}, \dots, A_N^{[\theta]} )$, i.e.,
\begin{align}
D(r)=\sum_{n=1}^N H\left(A_n^{[\theta]} \right).
\end{align}
In this work, we aim at characterizing the optimal normalized download cost $D^*(r)$ corresponding to every caching ratio $0 \leq r\leq 1$, where
\begin{align}
D^*(r)=\inf \left\{ \frac{D(r)}{L}: \left(D(r), L \right) \text{ is achievable}  \right\},
\end{align}
which is a function of the caching ratio $r$.

\section{Main Results}

We provide a PIR scheme for general $K$, $N$ and $r$, which achieves the following normalized download cost, $\bar{D}(r)$.
\begin{theorem}[Outer bound]\label{Thm1}
In the cache-aided PIR with partially known uncoded prefetching, let $s \in \{1,2, \cdots, K-1\}$, for the caching ratio $r_s$, where
\begin{align} \label{r_exp}
r_s=\frac{\binom{K-2}{s-1}}{\binom{K-2}{s-1}+\sum_{i=0}^{K-1-s} \binom{K-1}{s+i}(N-1)^{i+1}},
\end{align}
the optimal normalized download cost $D^*(r_s)$ is upper bounded by,
\begin{align} \label{eq_outer}
D^*(r_s) \leq \bar{D}(r_s) =\frac{\sum_{i=0}^{K-1-s} \binom{K}{s+1+i}(N-1)^{i+1} }{\binom{K-2}{s-1}+\sum_{i=0}^{K-1-s} \binom{K-1}{s+i}(N-1)^{i+1}}.
\end{align}
Moreover, if $r_s < r < r_{s+1}$, and $\alpha \in (0,1)$ such that $r=\alpha r_s+(1-\alpha) r_{s+1}$, then
\begin{align}
D^*(r) \leq \bar{D}(r) =\alpha \bar{D}(r_s)+(1-\alpha)\bar{D}(r_{s+1}).
\end{align}
\end{theorem}

The proof of Theorem~\ref{Thm1} is provided in Section~\ref{achievability}. The outer bound in Theorem~\ref{Thm1} is a piece-wise linear curve, which consists of $K$ line segments. These $K$ line segments intersect at the points $r_s$.

We characterize an inner bound (converse bound), which is denoted by $\tilde{D}(r)$, for the optimal normalized download cost $D^*(r)$ for general $K$, $N$, $r$.
\begin{theorem}[Inner bound]\label{Thm2}
In the cache-aided PIR with partially known uncoded prefetching, the normalized download cost is lower bounded as,
\begin{align} \label{eq_inner}
D^*(r) \geq \tilde{D}(r) &=\max_{i \in \{2, \cdots, K+1\}}  (1-r) \sum_{j=0} ^ {K+1-i} \frac{1}{N^j}-r\left(1-\frac{1}{N}\right)  \sum_{j=0}^{K-i} \frac{K+1-i-j}{N^j} \\
&=\max_{i \in \{2, \cdots, K+1\}} \sum_{j=0} ^ {K+1-i} \frac{1}{N^j}-(K+2-i)r.
\end{align}
\end{theorem}

The proof of Theorem~\ref{Thm2} is provided in Section~\ref{converse}. The inner bound in Theorem~\ref{Thm2} is also a piece-wise linear curve, which consists of $K$ line segments. Interestingly, these $K$ line segments intersect at the points as follows,
\begin{align} \label{tilde_ri}
\tilde{r}_i=\frac{1}{N^{K-i}}, \quad i=1, \cdots, K-1.
\end{align}

The outer bounds provided in Theorem~\ref{Thm1} and the inner bounds provided in Theorem~\ref{Thm2} match for some caching ratios $r$ as summarized in the following corollary.

\begin{corollary} [Optimal tradeoff for very low and very high caching ratios]\label{corollarylowhigh} ~
In the cache-aided PIR with partially known uncoded prefetching, for very low caching ratios, i.e., for $r \leq \frac{1}{ N^{K-1}}$, the optimal normalized download cost is given by,
\begin{align}\label{corollary2eqn}
D^*(r)&=\left(1+\frac{1}{N}+\cdots+\frac{1}{N^{K-1}}\right)-Kr.
\end{align}
On the other hand, for very high caching ratios, i.e., for $r \geq \frac{K-2}{N^2-3N+KN}$, the optimal normalized download cost is given by,
\begin{align} \label{corollary2eqn2}
D^*(r) =
\left\{
\begin{array}{ll}
1+\frac{1}{N}-2r, &\qquad \frac{K-2}{N^2-3N+KN} \leq r \leq \frac{1}{N} \\
1-r, &\qquad \frac{1}{N} \leq r  \leq 1
\end{array}
\right..
\end{align}
\end{corollary}	

\begin{Proof}
From \eqref{r_exp} and \eqref{tilde_ri}, we have
\begin{align}
r_1=\tilde{r}_1=& \frac{1}{ N^{K-1}},  \label{eq_tmp1} \\
r_{K-2}=&\frac{K-2}{N^2-3N+KN},  \label{eq_tmp2}       \\
r_{K-1}=\tilde{r}_{K-1}=&\frac{1}{N}. \label{eq_tmp3}
\end{align}

For the outer bound of the case of very low caching ratios, from \eqref{eq_outer}, we have
\begin{align}
\bar{D}(r_1)
&= \frac{\sum_{i=0}^{K-2} \binom{K}{2+i}(N-1)^{i+1}}{ \binom{K-2}{0}+\sum_{i=0}^{K-2} \binom{K-1}{1+i}(N-1)^{i+1}} \\
&= \frac{\frac{1}{(N-1)} \sum_{i=0}^{K-2}  \binom{K}{2+i}(N-1)^{i+2}}{ N^{K-1} } \\
&= \frac{\frac{1}{(N-1)} \left( N^K - 1 - K(N-1)   \right) }{N^{K-1}} \\
&= \frac{N^K-KN+K-1}{N^K-N^{K-1}}.
\end{align}	
For the inner bound of the case of very low caching ratios, from \eqref{eq_inner}, by choosing $i=2$ and using $r=r_1$, we have
\begin{align}
\tilde{D}(r_1)
&\geq (1-r_1) \sum_{j=0} ^ {K-1} \frac{1}{N^j} -r_1\left(1-\frac{1}{N}\right)  \sum_{j=0}^{K-2} \frac{K-1-j}{N^j} \\
&= \left( 1- \frac{1}{N^{K-1}} \right) \frac{1-\frac{1}{N^K} }{1-\frac{1}{N}}
- \frac{1}{N^{K-1}} \left(1-\frac{1}{N}\right) \frac{K-\frac{K}{N}-1+\frac{1}{N^K}}{\left(1-\frac{1}{N}\right) ^2 } \\
&= \frac{1}{\left(1-\frac{1}{N}\right)} \left[ \left( 1- \frac{1}{N^{K-1}} \right) \left ( 1-\frac{1}{N^K} \right)
- \frac{1}{N^{K-1}} \left( K-\frac{K}{N}-1+\frac{1}{N^K}  \right)     \right] \\
&= \frac{N^K-KN+K-1}{N^K-N^{K-1}}=\bar{D}(r_1).  \label{tmp1}
\end{align}
Thus, since $\tilde{D}(r_1) \leq \bar{D}(r_1)$ by definition, \eqref{tmp1} implies $\tilde{D}(r_1) = \bar{D}(r_1)$.

For the outer bound of the case of very high caching ratios, from \eqref{eq_outer}, we have
\begin{align}
\bar{D}(r_{K-2})&=\frac{\sum_{i=0}^{1} \binom{K}{K-1+i}(N-1)^{i+1} N}{ N \binom{K-2}{K-3}+\sum_{i=0}^{1}\binom{K-1}{K-2+i}(N-1)^{i+1}N} \\
&=\frac{N^2+KN-2N-K+1}{N^2-3N+KN},
\end{align}
and for the inner bound of the case of very high caching ratios, from \eqref{eq_inner} by choosing $i=K$ and using $r=r_{K-2}$,
\begin{align}
\tilde{D}(r_{K-2})&\geq (1-r_{K-2}) \sum_{j=0}^{1} \frac{1}{N^j} - r_{K-2} \left(1-\frac{1}{N}\right) \sum_{j=0}^{0} \frac{1-j}{N^j} \\
&= 1+\frac{1}{N} - 2r_{K-2}  \\
&= \frac{N^2+KN-2N-K+1}{N^2-3N+KN}   =  \bar{D}(r_{K-2}) \label{tmp2}
\end{align}
implying $\tilde{D}(r_{K-2})=\bar{D}(r_{K-2})$.

Finally, from \eqref{eq_outer}, $\bar{D}(r_{K-1})=\frac{N-1}{N}$, and from \eqref{eq_inner} by choosing $i=K+1$ and using $r=r_{K-1}$,
\begin{align}
\tilde{D}(r_{K-1})\geq\frac{N-1}{N}=\bar{D}(r_{K-1}) \label{tmp3}
\end{align}
implying $\tilde{D}(r_{K-1})=\bar{D}(r_{K-1})$.

Therefore, $\tilde{D}(r)=\bar{D}(r)$ at $r=r_1$, $r=r_{K-2}$ and $r=r_{K-1}$. In addition to that $\tilde{D}(0)=\bar{D}(0)$ and $\tilde{D}(1)=\bar{D}(1)$. Since both $\bar{D}(r)$ and $\tilde{D}(r)$ are linear functions of $r$, and since $\tilde{D}(0)=\bar{D}(0)$ and $\tilde{D}(r_1)=\bar{D}(r_1)$, we have $\tilde{D}(r)=\bar{D}(r)=D^*(r)$ for $0\leq r\leq r_1$. This is the very low caching ratio region. In addition, since $\tilde{D}(r_{K-2})=\bar{D}(r_{K-2})$, $\tilde{D}(r_{K-1})=\bar{D}(r_{K-1})$ and $\tilde{D}(1)=\bar{D}(1)$, we have $\tilde{D}(r)=\bar{D}(r)=D^*(r)$ for $r_{K-2}\leq r\leq 1$. This is the very high caching ratio region.
\end{Proof}	

We use the example of $K=4$, $N=2$ to illustrate Corollary~\ref{corollarylowhigh} (see Figure~\ref{K=4,N=2 case}). In this case, $r_1=\tilde{r}_1=\frac{1}{8}$, $r_{K-2}=\frac{1}{3}$, and $r_{K-1}=\tilde{r}_{K-1}=\frac{1}{2}$. Therefore, we have exact results for $0\leq r \leq \frac{1}{8}$ (very low caching ratios) and $\frac{1}{3}\leq r \leq 1$ (very high caching ratios). We have a gap between the achievability and the converse for medium caching ratios in $\frac{1}{8}\leq r \leq \frac{1}{3}$. More specifically, line segments connecting $(0,\frac{15}{8})$ and $(\frac{1}{8},\frac{11}{8})$; connecting $(\frac{1}{3},\frac{5}{6})$ and $(\frac{1}{2},\frac{1}{2})$; and connecting $(\frac{1}{2},\frac{1}{2})$ and $(1,0)$ are tight.
\begin{figure}[t]
\centering
\epsfig{file=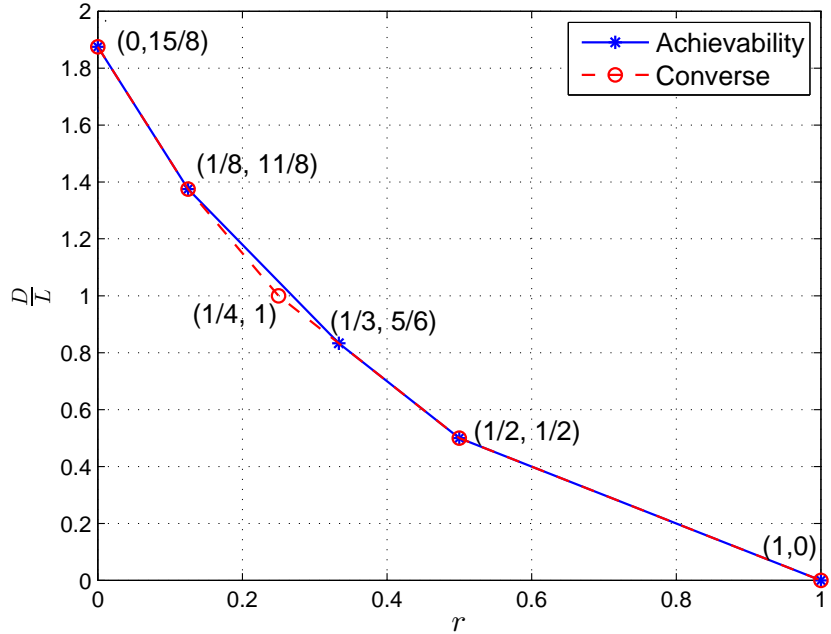,width=0.75\textwidth}
\caption{Inner and outer bounds for $K=4$, $N=2$.}
\label{K=4,N=2 case}
\end{figure}

For the case $K=3$, we have exact tradeoff curve for any $N$, $r$ as shown in the following corollary.
\begin{corollary}[Optimal tradeoff for $K=3$]\label{corollaryK=3}
In the cache-aided PIR with partially known uncoded prefetching with $K=3$ messages, the optimal download cost caching ratio tradeoff is given explicitly as,
\begin{equation} \label{tmp4}
D^*(r) =
\left\{
\begin{array}{ll}
1+\frac{1}{N}+\frac{1}{N^2}-3r,
&\qquad 0\leq r\leq \frac{1}{N^2} \\
1+\frac{1}{N}-2r,  &\qquad\frac{1}{N^2} \leq r \leq \frac{1}{N} \\
1-r, &\qquad\frac{1}{N} \leq r  \leq 1
\end{array}
\right..
\end{equation}
\end{corollary}
\begin{Proof}
The proof follows from the proof of Corollary~\ref{corollarylowhigh}. Note that in this case, from \eqref{eq_tmp1} and \eqref{eq_tmp2}, $r_1=r_{K-2}=\frac{1}{N^2}$; and from \eqref{eq_tmp3}, $r_2=r_{K-1}=\frac{1}{N}$. Thus, we have a tight result for $0\leq r \leq r_1=\frac{1}{N^2}$ (very low caching ratios) and a tight result for $r_{K-2}=r_1=\frac{1}{N^2}\leq r \leq 1$, i.e., a tight result for all $0\leq r \leq 1$. We have three segments in this case: $[0, r_1]$, $[r_1, r_2]$ and $[r_2,1]$ with three different line expressions for the exact result as given in \eqref{r_exp}-\eqref{eq_outer} and written explicitly in \eqref{tmp4}.
\end{Proof}

\section{Achievable Scheme}\label{achievability}

In this section, we present an achievable scheme for the outer bounds provided in Theorem~\ref{Thm1}. Our achievable scheme is based on \cite{JafarPIR, tandon2017capacity, wei2017fundamental}. We first provide achievable schemes for the caching ratios $r_s$ in \eqref{r_exp} by applying the principles in \cite{JafarPIR}: 1) database symmetry, 2) message symmetry within each database, and 3) exploiting undesired messages as side information. For an arbitrary caching ratio $r \neq r_s$, we apply the memory-sharing scheme in \cite{tandon2017capacity}. Since the cached content is partially known by the databases, the achievable scheme is different from that in \cite{wei2017fundamental}. We first use the case of $K=3$, $N=2$ to illustrate the main ideas of our achievability scheme.

\subsection{Motivating Example: $K=3$ Messages and $N=2$ Databases}

We use $a_i$, $b_i$, and $c_i$ to denote the bits of messages $W_1$, $W_2$ and $W_3$, respectively. We assume that the user wants to retrieve message $W_1$ privately without loss of generality.

\subsubsection{Caching Ratio $r_1=\frac{1}{4}$} \label{r=1/4}

We choose the message size as $8$ bits. In the prefetching phase, for caching ratio $r_1=\frac{1}{4}$, the user caches $2$ bits from each message. Therefore, the user caches $1$ bit from each database for each message. Therefore, $Z_1=(a_1, b_1, c_1)$ and $Z_2=(a_2, b_2, c_2)$.

In the retrieval phase, for $s=1$, we first mix $1$ bit of side information with the desired bit. Therefore, the user queries $a_3+b_2$ and $a_4+c_2$ from database $1$. Note that database $1$ knows that the user prefetches $Z_1$. Therefore, the user does not use side information $Z_1$ to retrieve information from database $1$. To keep message symmetry, the user further queries $b_3+c_3$ from database $1$. Similarly, the user queries $a_5+b_1$, $a_6+c_1$ and $b_4+c_4$ from database $2$. Then, the user exploits the side information $b_4+c_4$ to query $a_7+b_4+c_4$ from database $1$ and the side information $b_3+c_3$ to query $a_8+b_3+c_3$ from database $2$. After this step, no more side information can be used and the message symmetry is attained for each database. Therefore, the PIR scheme ends here. The decodability of message $W_1$ can be shown easily, since the desired bits are either mixed with cached side information or the side information obtained from other databases. Overall, the user downloads $8$ bits. Therefore, the normalized download cost is $1$. We summarize the queries in Table.~\ref{K=3 r=1/4}.

\begin{table}[H]
\centering
\caption{Query table for $K=3$, $N=2$, $r_1=\frac{1}{4}$.}
\label{K=3 r=1/4}
\begin{tabular}{ccc}
\hline
\multicolumn{1}{|c|}{$s$}                    & \multicolumn{1}{c|}{DB1}           & \multicolumn{1}{c|}{DB2}            \\ \hline
\multicolumn{1}{|c|}{\multirow{2}{*}{\rotatebox[origin=c]{90}{\parbox[c]{1cm}{\centering $s=1$}}}}
                                             & \multicolumn{1}{c|}{$a_3+b_2$}     & \multicolumn{1}{c|}{$a_5+b_1$}      \\ \cline{2-3}
\multicolumn{1}{|c|}{}                       & \multicolumn{1}{c|}{$a_4+c_2$}     & \multicolumn{1}{c|}{$a_6+c_1$}      \\ \hline
\multicolumn{1}{|c|}{\multirow{2}{*}{}}      & \multicolumn{1}{c|}{$b_3+c_3$}     & \multicolumn{1}{c|}{$b_4+c_4$}      \\ \cline{2-3}
\multicolumn{1}{|c|}{}                       & \multicolumn{1}{c|}{$a_7+b_4+c_4$} & \multicolumn{1}{c|}{$a_8+b_3+c_3$}  \\ \hline
	 	               &                                        &                                                       \\ \cline{2-3}
\multicolumn{1}{c|}{}  &\multicolumn{1}{c|}{$Z_1=(a_1,b_1,c_1)$}&\multicolumn{1}{c|}{$Z_2=(a_2,b_2,c_2)$}               \\ \cline{2-3}
\end{tabular}
\end{table}

\subsubsection{Caching Ratio $r_2=\frac{1}{2}$} \label{r=1/2}

We choose the message size as $4$ bits. In the prefetching phase, for caching ratio $r_2=\frac{1}{2}$, the user caches $2$ bits from each message. Therefore, the user caches $1$ bit from each database for each message. Therefore, $Z_1=(a_1, b_1, c_1)$ and $Z_2=(a_2, b_2, c_2)$. In the retrieval phase, for $s=2$, we first mix $2$ bits of side information with the desired bit. Therefore, the user queries $a_3+b_2+c_2$ from database $1$. Similarly, the user queries $a_4+b_1+c_1$ from database $2$. After this, no more side information can be used and the message symmetry is attained for each database. Therefore, the PIR scheme ends here. Overall, the user downloads $2$ bits. Therefore, the normalized download cost is $\frac{1}{2}$. We summarize the queries in Table.~\ref{K=3 r=1/2}.

\begin{table}[H]
\centering
\caption{Query table for $K=3$, $N=2$, $r_2=\frac{1}{2}$.}
\label{K=3 r=1/2}
\begin{tabular}{ccc}
\hline
\multicolumn{1}{|c|}{$s$}       & \multicolumn{1}{c|}{DB1}           & \multicolumn{1}{c|}{DB2}            \\ \hline
\multicolumn{1}{|c|}{$s=2$}  	& \multicolumn{1}{c|}{$a_3+b_2+c_2$}     & \multicolumn{1}{c|}{$a_4+b_1+c_1$}      \\ \hline
		&                                        &                                                       \\ \cline{2-3}
		\multicolumn{1}{c|}{}  &\multicolumn{1}{c|}{$Z_1=(a_1,b_1,c_1)$}&\multicolumn{1}{c|}{$Z_2=(a_2,b_2,c_2)$}               \\ \cline{2-3}
	\end{tabular}
\end{table}

\subsubsection{Caching Ratio $r=\frac{1}{3}$} \label{r=1/3}

We choose the message size as $12$ bits. In the prefetching phase, for caching ratio $r=\frac{1}{3}$, the user caches $4$ bits from each message. Therefore, the user caches $2$ bits from each database for each message. Therefore, $Z_1=(a_1,a_2,b_1,b_2, c_1, c_2)$ and $Z_2=(a_3,a_4,b_3,b_4, c_3, c_4)$. In the retrieval phase, we combine the achievable schemes in Section~\ref{r=1/4} and \ref{r=1/2} as shown in Table~\ref{K=3 r=1/3}. The normalized download cost is $\frac{5}{6}$. By applying \cite[Lemma~1]{tandon2017capacity} and taking $\alpha=\frac{2}{3}$, we can show that $\bar{D}(\frac{1}{3})=\bar{D}(\frac{2}{3} \cdot \frac{1}{4} + \frac{1}{3}\cdot \frac{1}{2} )=\frac{2}{3}\bar{D}(\frac{1}{4})+\frac{1}{3}\bar{D}(\frac{1}{2})=\frac{2}{3}\cdot 1+\frac{1}{3}\cdot\frac{1}{2}=\frac{5}{6}$.

\begin{table}[H]
\centering
\caption{Query table for $K=3$, $N=2$, $r=\frac{1}{3}$.}
\label{K=3 r=1/3}
\begin{tabular}{ccc}
\hline
\multicolumn{1}{|c|}{$s$}       & \multicolumn{1}{c|}{DB1}           & \multicolumn{1}{c|}{DB2}            \\ \hline
\multicolumn{1}{|c|}{\multirow{2}{*}{\rotatebox[origin=c]{90}{\parbox[c]{1cm}{\centering $s=1$}}}}
& \multicolumn{1}{c|}{$a_5+b_3$}     & \multicolumn{1}{c|}{$a_7+b_1$}      \\ \cline{2-3}
\multicolumn{1}{|c|}{}                       & \multicolumn{1}{c|}{$a_6+c_3$}     & \multicolumn{1}{c|}{$a_8+c_1$}      \\ \hline
\multicolumn{1}{|c|}{\multirow{2}{*}{}}      & \multicolumn{1}{c|}{$b_5+c_5$}     & \multicolumn{1}{c|}{$b_6+c_6$}      \\ \cline{2-3}
\multicolumn{1}{|c|}{}                       & \multicolumn{1}{c|}{$a_9+b_6+c_6$} & \multicolumn{1}{c|}{$a_{10}+b_5+c_5$}  \\ \hline
\multicolumn{1}{|c|}{$s=2$}  	& \multicolumn{1}{c|}{$a_{11}+b_4+c_4$}     & \multicolumn{1}{c|}{$a_{12}+b_2+c_2$}      \\ \hline
&                                        &                                                       \\ \cline{2-3}
\multicolumn{1}{c|}{}  &\multicolumn{1}{c|}{$Z_1=(a_1,a_2,b_1,b_2, c_1, c_2)$}
                       &\multicolumn{1}{c|}{$Z_2=(a_3,a_4,b_3,b_4, c_3, c_4)$}               \\ \cline{2-3}
\end{tabular}
\end{table}

\subsection{Achievable Scheme}
We first present the achievable scheme for the caching ratios $r_s$ given in \eqref{r_exp}. Then, we apply the memory-sharing scheme provided in \cite{tandon2017capacity} for the intermediate caching ratios.

\subsubsection{Achievable Scheme for the Caching Ratio $r_s$}
For fixed $K$ and $N$, there are $K-1$ non-degenerate corner points (in addition to degenerate caching ratios $r=0$ and $r=1$). The caching ratios, $r_s$, corresponding to these non-degenerate corner points are indexed by $s$, which represents the number of cached bits used in the side information mixture at the first round of the querying. For each $s\in\{1,2,\dots, K-1\}$, we choose the length of the message to be $L(s)$ for the corner point indexed by $s$, where
\begin{align}
L(s)=N\binom{K-2}{s-1}+\sum_{i=0}^{K-1-s} \binom{K-1}{s+i}(N-1)^{i+1}N.
\end{align}

In the prefetching phase, for each message the user randomly and independently chooses $N\binom{K-2}{s-1}$ bits to cache, and caches $\binom{K-2}{s-1}$ bits from each database for each message. Therefore, the caching ratio $r_s$ is equal to
\begin{align} \label{r_exact}
r_s=\frac{N\binom{K-2}{s-1}}{N\binom{K-2}{s-1}+\sum_{i=0}^{K-1-s} \binom{K-1}{s+i}(N-1)^{i+1}N}.
\end{align}

In the retrieval phase, the user applies the PIR scheme as follows:
\begin{enumerate}
\item \emph{Initialization:} Set the round index to $t=s+1$, where the $t$th round involves downloading sums of every $t$ combinations of the $K$ messages.
\item \emph{Exploiting side information:} If $t=s+1$, for the first database, the user forms queries by mixing $s$ undesired bits cached from the other $N-1$ databases in the prefetching phase to form one side information equation. Each side information equation is added to one bit from the uncached portion of the desired message. Therefore, for the first database, the user downloads $\binom{K-1}{s}(N-1)$ equations in the form of a desired bit added to a mixture of $s$ cached bits from other messages. On the other hand, if $t>s+1$, for the first database, the user exploits the $\binom{K-1}{t-1}(N-1)^{t-s}$ side information equations generated from the remaining $(N-1)$ databases in the $(t-1)$th round.
\item \emph{Symmetry across databases:} The user downloads the same number of equations with the same structure as in step $2$ from every database. Consequently, the user decodes $\binom{K-1}{t-1}(N-1)^{t-s}$ desired bits from every database, which are done either using the cached bits as side information if $t=s+1$, or the side information generated in the $(t-1)$th round if $t>s+1$.
\item \emph{Message symmetry:} To satisfy the privacy constraint, the user should download the same amount of bits from other messages. Therefore, the user downloads $\binom{K-1}{t}(N-1)^{t-s}$ undesired equations from each database in the form of sum of $t$ bits from the uncached portion of the undesired messages.
\item \emph{Repeat} steps 2, 3, 4 after setting $t=t+1$ until $t=K$.
\item \emph{Shuffling the order of queries:} By shuffling the order of queries uniformly, all possible queries can be made equally likely regardless of the message index.
\end{enumerate}

Since the desired bits are added to the side information which is either obtained from the cached bits (if $t=s+1$) or from the remaining $(N-1)$ databases in the $(t-1)$th round when $t>s+1$, the user can decode the uncached portion of the desired message by canceling out the side information bits. In addition, for each database, each message is queried equally likely with the same set of equations, which guarantees privacy as in \cite{JafarPIR}. Therefore, the privacy constraint in \eqref{privacy_constraint} and the reliability constraint in \eqref{reliability_constraint} are satisfied.

We now calculate the total number of downloaded bits for the caching ratio $r_s$ in \eqref{r_exact}. For the round $t=s+1$, we exploit $s$ cached bits to form the side information equation. Therefore, each download is a sum of $s+1$ bits. For each database, we utilize the side information cached from other $N-1$ databases. In addition to the message symmetry step enforcing symmetry across $K$ messages, we download $\binom{K}{s+1}(N-1)$ bits from a database. Due to the database symmetry step, in total, we download $\binom{K}{s+1}(N-1)N$ bits. For the round $t=s+i>s+1$, we exploit $s+i-1$ undesired bits downloaded from the $(t-1)$th round to form the side information equation. Due to message symmetry and database symmetry, we download $\binom{K}{s+1+i} (N-1)^{i+1} N$ bits. Overall, the total number of downloaded bits is,
\begin{align}
D(r_s)= \sum_{i=0}^{K-1-s} \binom{K}{s+1+i} (N-1)^{i+1} N.
\end{align}
By canceling out the undesired side information bits using the cached bits for the round $t=s+1$, we obtain $\binom{K-1}{s}(N-1)N$ desired bits. For the round $t=s+i>s+1$, we decode $\binom{K-1}{s+i}(N-1)^{i+1}N$ desired bits by using the side information obtained in $(t-1)$th round. Overall, we obtain $L(s)-N\binom{K-2}{s-1}$ desired bits. Therefore, the normalized download cost is,
\begin{align} \label{D(r_s)}
\bar{D}(r_s)=\frac{D(r_s)}{L(s)}=\frac{\sum_{i=0}^{K-1-s} \binom{K}{s+1+i}(N-1)^{i+1} N}{N\binom{K-2}{s-1}+\sum_{i=0}^{K-1-s} \binom{K-1}{s+i}(N-1)^{i+1}N}.
\end{align}

\subsubsection{Achievable Scheme for the Caching Ratios not Equal to $r_s$}

For caching ratios $r$ which are not exactly equal to \eqref{r_exact} for some $s$, we first find an $s$ such that $r_s<r<r_{s+1}$. We choose $0<\alpha<1$ such that $r=\alpha r_s + (1-\alpha) r_{s+1}$. By using the memory-sharing scheme in \cite[Lemma~1]{tandon2017capacity}, we achieve the following normalized download cost,
\begin{align}
\bar{D}(r) = \alpha \bar{D}(r_s)+(1-\alpha) \bar{D}(r_{s+1}).
\end{align}

\section{Converse Proof} \label{converse}

In this section, we derive an inner bound for the cache-aided PIR with partially known uncoded prefetching. We extend the techniques in \cite{JafarPIR, wei2017fundamental} to our problem. The main difference between this proof and that in \cite{wei2017fundamental} is the usage of privacy constraint given in \eqref{privacy_constraint}.

\begin{lemma}[Interference lower bound]\cite[Lemma~1]{wei2017fundamental}\label{lemma_converse1}
For the cache-aided PIR with partially known uncoded prefetching, the interference from undesired messages within the answering strings $D(r)-L(1-r)$ is lower bounded by, 	
\begin{align}
D(r) -L(1-r) + o(L) \geq I\left(W_{k:K}; \mathbb{H}, Q_{1:N}^{[k-1]}, A_{1:N}^{[k-1]}|W_{1:k-1}, Z \right) \label{eq_L1}
\end{align}
for all $k\in \{2,\dots,K\}$.
\end{lemma}

The proof of Lemma~\ref{lemma_converse1} is similar to \cite[Lemma~1]{wei2017fundamental}. In the following lemma, we prove an inductive relation for the mutual information term on the right hand side of \eqref{eq_L1}.
\begin{lemma}[Induction lemma]\label{lemma_converse2}
For all $k\in \{2,\dots,K\}$, the mutual information term in Lemma~\ref{lemma_converse1} can be inductively lower bounded as,
\begin{align} \label{eq_L2}
&I\left( W_{k:K} ; \mathbb{H},  Q_{1:N}^{[k-1]}, A_{1:N}^{[k-1]}  | W_{1:k-1}, Z \right)  \notag \\
&\quad \geq \frac{1}{N}  I\left(W_{k+1:K};\mathbb{H},  Q_{1:N}^{[k]}, A_{1:N}^{[k]}|W_{1:k},  Z  \right) +\frac{L(1-r) - o(L)}{N}  +  \frac{1-N}{N}  (K-k+1)Lr.
\end{align}
\end{lemma}

Lemma~\ref{lemma_converse2} is a generalization of \cite[Lemma~6]{JafarPIR} and \cite[Lemma~2]{wei2017fundamental}, and it reduces to \cite[Lemma~6]{JafarPIR} when $r=0$. Compared to \cite[Lemma~2]{wei2017fundamental}, the lower bound in \eqref{eq_L2} is increased by $\frac{(K-k+1)Lr}{N}$, since the cached content is partially known by the databases.

\begin{Proof}
We start with the left hand side of \eqref{eq_L2},
\begin{align}
&I\left( W_{k:K} ; \mathbb{H}, Q_{1:N}^{[k-1]}, A_{1:N}^{[k-1]} | W_{1:k-1}, Z \right)       \notag \\
&\qquad  =  I\left(W_{k:K} ; \mathbb{H}, Q_{1:N}^{[k-1]}, A_{1:N}^{[k-1]}, Z | W_{1:k-1}  \right)  - I(W_{k:K};Z|W_{1:k-1}).    \label{eq_L2_a}
\end{align}

For the first term on the right hand side of \eqref{eq_L2_a}, we have
\begin{align}
&I\left(W_{k:K} ; \mathbb{H}, Q_{1:N}^{[k-1]}, A_{1:N}^{[k-1]}, Z| W_{1:k-1}  \right) \notag \\
&\quad= \frac{1}{N}NI\left(W_{k:K} ; \mathbb{H}, Q_{1:N}^{[k-1]}, A_{1:N}^{[k-1]}, Z| W_{1:k-1}  \right)  \\
&\label{eq_IL_1}\quad \geq\frac{1}{N}\sum_{n=1}^N I\left(W_{k:K} ; \mathbb{H}_n, Q_n^{[k-1]}, A_n^{[k-1]}, Z_n| W_{1:k-1}  \right) \\
&\quad = \frac{1}{N} \left[\sum_{n=1}^N I\left(W_{k:K};\mathbb{H}_n,Z_n| W_{1:k-1}  \right) +
          \sum_{n=1}^N I\left(W_{k:K};Q_n^{[k-1]}, A_n^{[k-1]}|W_{1:k-1},\mathbb{H}_n,Z_n\right)  \right] \\
&\label{eq_IL_2}\quad \geq \frac{1}{N} \left[\sum_{n=1}^N I\left(W_{k:K};Z_n| W_{1:k-1}  \right) +
\sum_{n=1}^N I\left(W_{k:K};Q_n^{[k-1]}, A_n^{[k-1]}|W_{1:k-1},\mathbb{H}_n,Z_n\right)  \right] \\
&\label{eq_IL_3}\quad = \frac{1}{N} \left[ N \times \frac{(K-k+1)Lr}{N}
+\sum_{n=1}^N I\left(W_{k:K};Q_n^{[k-1]}, A_n^{[k-1]}|W_{1:k-1},\mathbb{H}_n,Z_n\right)    \right] \\
&\quad = \frac{(K-k+1)Lr}{N} + \frac{1}{N}\sum_{n=1}^N I\left(W_{k:K};Q_n^{[k-1]}, A_n^{[k-1]}|W_{1:k-1},\mathbb{H}_n,Z_n\right)  \\
&\label{eq_IL_4}\quad \stackrel{\eqref{privacy_constraint}}{=} \frac{(K-k+1)Lr}{N} + \frac{1}{N}\sum_{n=1}^N I\left(W_{k:K};Q_n^{[k]}, A_n^{[k]}|W_{1:k-1},\mathbb{H}_n,Z_n\right)  \\
&\label{eq_IL_5}~ \stackrel{\eqref{independency},\eqref{query_indep}}{=} \frac{(K-k+1)Lr}{N} + \frac{1}{N}\sum_{n=1}^N I\left(W_{k:K};A_n^{[k]}|W_{1:k-1},\mathbb{H}_n,Z_n, Q_n^{[k]}\right)  \\
&\label{eq_IL_7}\quad \stackrel{\eqref{answer_constraint}}{=} \frac{(K-k+1)Lr}{N} + \frac{1}{N}\sum_{n=1}^N H\left(A_n^{[k]}|W_{1:k-1},\mathbb{H}_n ,Z_n, Q_n^{[k]}\right)  \\
&\label{eq_IL_9}\quad \geq \frac{(K-k+1)Lr}{N} +  \frac{1}{N}\sum_{n=1}^N H\left(A_n^{[k]}|W_{1:k-1},\mathbb{H},Z, Q_{1:N}^{[k]}, A_{1:n-1}^{[k]}\right)  \\
&\label{eq_IL_8}\quad \stackrel{\eqref{answer_constraint}}{=} \frac{(K-k+1)Lr}{N}  + \frac{1}{N}\sum_{n=1}^N I \left(W_{k:K};A_n^{[k]}|W_{1:k-1},\mathbb{H},Z, Q_{1:N}^{[k]}, A_{1:n-1}^{[k]}   \right) \\
&\quad = \frac{(K-k+1)Lr}{N}  + \frac{1}{N} I \left(W_{k:K};A_{1:N}^{[k]}|W_{1:k-1},\mathbb{H},Z, Q_{1:N}^{[k]}   \right) \\
&\label{eq_IL_6}~ \stackrel{\eqref{independency},\eqref{query_indep}}{=} \frac{(K-k+1)Lr}{N} +
\frac{1}{N} I \left(W_{k:K};\mathbb{H},Q_{1:N}^{[k]}, A_{1:N}^{[k]}|W_{1:k-1},Z    \right) \\
&\label{eq_IL_10}\quad \stackrel{\eqref{reliability_constraint}}{=} \frac{(K-k+1)Lr}{N} +
\frac{1}{N} I \left(W_{k:K};W_k, \mathbb{H},Q_{1:N}^{[k]}, A_{1:N}^{[k]}|W_{1:k-1},Z \right) - \frac{o(L)}{N} \\
&\quad = \frac{(K-k+1)Lr}{N} + \frac{1}{N} \left[I \left(W_{k:K};W_k|W_{1:k-1},Z \right) +
I \left(W_{k:K}; \mathbb{H},Q_{1:N}^{[k]}, A_{1:N}^{[k]}|W_{1:k},Z \right)  \right]\notag \\
&\qquad  - \frac{o(L)}{N} \\
&\quad = \frac{(K-k+1)Lr}{N} + \frac{L(1-r)}{N} + \frac{1}{N} I \left(W_{k+1:K}; \mathbb{H},Q_{1:N}^{[k]}, A_{1:N}^{[k]}|W_{1:k},Z \right) - \frac{o(L)}{N},  \label{eq_L2_b}
\end{align}
where \eqref{eq_IL_1} and \eqref{eq_IL_2} follow from the non-negativity of mutual information, \eqref{eq_IL_3} is due to the fact that from the $n$th database, the user prefetches $\frac{KLr}{N}$ bits, \eqref{eq_IL_4} follows from the privacy constraint, \eqref{eq_IL_5} and \eqref{eq_IL_6} follow from the independence of $W_{k:K}$ and $Q_n^{[k]}$, \eqref{eq_IL_7} and \eqref{eq_IL_8} follow from the fact that the answering string $A_n^{[k]}$ is a deterministic function of $(W_{1:K}, Q_n^{[k]})$, \eqref{eq_IL_9} follows from conditioning reduces entropy, and \eqref{eq_IL_10} follows from the reliability constraint.

For the second term on the right hand side of \eqref{eq_L2_a}, we have
\begin{align}
I(W_{k:K};Z|W_{1:k-1})&= H\left(Z|W_{1:k-1}\right)-H(Z|W_{1:K}) \\
&=\left( K-k+1 \right) L r  \label{eq_L2_c}
\end{align}	
where \eqref{eq_L2_c} follows from the uncoded nature of the cached bits.
	
Combining \eqref{eq_L2_a}, \eqref{eq_L2_b} and \eqref{eq_L2_c} yields \eqref{eq_L2}.
\end{Proof}

Now, we are ready to derive the general inner bound for arbitrary $K$, $N$, $r$. To obtain this bound, we use Lemma~\ref{lemma_converse1} to find $K$ lower bounds by varying the index $k$ in the lemma from $k=2$ to $k=K$, and by using the non-negativity of mutual information for the $K$th bound. Next, we inductively lower bound each term of Lemma~\ref{lemma_converse1} by using Lemma~\ref{lemma_converse2} $(K-k+1)$ times to get $K$ explicit lower bounds.

\begin{lemma}
For fixed $N$, $K$ and $r$, we have
\begin{align}
D(r) \geq L(1-r) \sum_{j=0} ^ {K+1-k} \frac{1}{N^j} -Lr\left(1-\frac{1}{N}\right)  \sum_{j=0}^{K-k} \frac{K+1-k-j}{N^j} + o(L),
\end{align}
where $k=2, \dots, K+1$.
\end{lemma}
\begin{Proof}
We have
\begin{align}	
D(r)
&\label{L3_1}\stackrel{\eqref{eq_L1}}{\geq} L(1-r) +  I\left(W_{k:K}; \mathbb{H}, Q_{1:N}^{[k-1]}, A_{1:N}^{[k-1]}|W_{1:k-1}, Z \right) -o(L) \\
&\stackrel{\eqref{eq_L2}}{\geq} L(1-r) + \frac{L(1-r)}{N} - (K-k+1) Lr + \frac{1}{N} (K-k+1) Lr \notag \\
&\qquad + \frac{1}{N}  I\left(W_{k+1:K}; \mathbb{H}, Q_{1:N}^{[k]}, A_{1:N}^{[k]}|W_{1:k}, Z \right) -o(L)  \\
&\stackrel{\eqref{eq_L2}}{\geq} L(1-r) \left[1+\frac{1}{N}+\frac{1}{N^2}\right] +\left(\frac{Lr}{N}-Lr\right)   \left[(K-k+1)+\frac{(K-k)}{N}\right]
 \notag \\
&\qquad + \frac{1}{N^2}  I\left(W_{k+2:K}; \mathbb{H}, Q_{1:N}^{[k+1]}, A_{1:N}^{[k+1]}|W_{1:k+1}, Z \right) -o(L)   \\
&\stackrel{\eqref{eq_L2}}{\geq} \dots \\
&\stackrel{\eqref{eq_L2}}{\geq} L(1-r) \sum_{j=0} ^ {K+1-k} \frac{1}{N^j} -Lr\left(1-\frac{1}{N}\right)  \sum_{j=0}^{K-k} \frac{K+1-k-j}{N^j} + o(L),
\end{align}
where \eqref{L3_1} follows from Lemma~\ref{lemma_converse1}, and the remaining steps follow from the successive application of Lemma~\ref{lemma_converse2}.
\end{Proof}

We conclude the converse proof by dividing by $L$ and taking the limit as $L \rightarrow \infty$. Then, for $k=2, \cdots, K+1$, we have
\begin{align}\label{beforelast}
D^*(r) \geq (1-r) \sum_{j=0} ^ {K+1-k} \frac{1}{N^j} -r\left(1-\frac{1}{N}\right)  \sum_{j=0}^{K-k} \frac{K+1-k-j}{N^j}.
\end{align}
Since \eqref{beforelast} gives $K$ intersecting line segments, the normalized download cost is lower bounded by their maximum value as follows
\begin{align}
D^*(r) \geq \max_{i \in \{2, \cdots, K+1\}} (1-r) \sum_{j=0} ^ {K+1-i} \frac{1}{N^j} -r\left(1-\frac{1}{N}\right)  \sum_{j=0}^{K-i} \frac{K+1-i-j}{N^j}.
\end{align}

\section{Further Examples}

\subsection{$K=4$ Messages, $N=2$ Databases}

For $K=4$ and $N=2$, we present achievable PIR schemes for caching ratios $r_1=\frac{1}{8}$ in Table~\ref{N_2_K_4_r=1/8}, $r_2=\frac{1}{3}$ in Table~\ref{N_2_K_4_r=1/3}, and $r_3=\frac{1}{2}$ in Table~\ref{N_2_K_4_r=1/2}. The PIR schemes aim to retrieve message $W_1$, where we use $a_i$ to denote its bits. The achievable normalized download costs for these caching ratios are $\frac{11}{8}$, $\frac{5}{6}$ and $\frac{1}{2}$, respectively. The plot of the inner and outer bounds can be found in Figure~\ref{K=4,N=2 case}.
\begin{table}[H]
\centering
\caption{Query table for $K=4$, $N=2$ and $r_1=\frac{1}{8}$.}
\label{N_2_K_4_r=1/8}
\begin{tabular}{ccc}
\hline
\multicolumn{1}{|c|}{$s$}                    & \multicolumn{1}{c|}{DB1}           & \multicolumn{1}{c|}{DB2}           \\ \hline
\multicolumn{1}{|c|}{\multirow{3}{*}{\rotatebox[origin=c]{90}{\parbox[c]{1cm}{\centering $s=1$}}}}
                                             & \multicolumn{1}{c|}{$a_3+b_2$}     & \multicolumn{1}{c|}{$a_6+b_1$}     \\ \cline{2-3}
\multicolumn{1}{|c|}{}                       & \multicolumn{1}{c|}{$a_4+c_2$}     & \multicolumn{1}{c|}{$a_7+c_1$}     \\ \cline{2-3}
\multicolumn{1}{|c|}{}                       & \multicolumn{1}{c|}{$a_5+d_2$}     & \multicolumn{1}{c|}{$a_8+d_1$}     \\ \hline
\multicolumn{1}{|c|}{\multirow{2}{*}{}}      & \multicolumn{1}{c|}{$b_3+c_3$}     & \multicolumn{1}{c|}{$b_5+c_5$}     \\ \cline{2-3}
\multicolumn{1}{|c|}{}                       & \multicolumn{1}{c|}{$b_4+d_3$}     & \multicolumn{1}{c|}{$b_6+d_5$}      \\ \cline{2-3}
\multicolumn{1}{|c|}{}                       & \multicolumn{1}{c|}{$c_4+d_4$}     & \multicolumn{1}{c|}{$c_6+d_6$}      \\ \cline{2-3}
\multicolumn{1}{|c|}{}                       & \multicolumn{1}{c|}{$a_9+b_5+c_5$} & \multicolumn{1}{c|}{$a_{12}+b_3+c_3$} \\ \cline{2-3}
\multicolumn{1}{|c|}{}                   & \multicolumn{1}{c|}{$a_{10}+b_6+d_5$} & \multicolumn{1}{c|}{$a_{13}+b_4+d_3$} \\ \cline{2-3}
\multicolumn{1}{|c|}{}                   & \multicolumn{1}{c|}{$a_{11}+c_6+d_6$}&\multicolumn{1}{c|}{$a_{14}+c_4+d_4$} \\ \cline{2-3}
\multicolumn{1}{|c|}{}                   & \multicolumn{1}{c|}{$b_7+c_7+d_7$} & \multicolumn{1}{c|}{$b_8+c_8+d_8$} \\ \cline{2-3}
\multicolumn{1}{|c|}{}                & \multicolumn{1}{c|}{$a_{15}+b_8+c_8+d_8$} & \multicolumn{1}{c|}{$a_{16}+b_7+c_7+d_7$} \\ \hline
		&                                    &                                    \\ \cline{2-3}
\multicolumn{1}{c|}{}   &\multicolumn{1}{c|}{$Z_1=(a_1,b_1,c_1,d_1)$}&\multicolumn{1}{c|}{$Z_2=(a_2,b_2,c_2,d_2)$}        \\ \cline{2-3}
\end{tabular}
\end{table}

\begin{table}[H]
\centering
\caption{Query table for $K=4$, $N=2$, $r_2=\frac{1}{3}$.}
\label{N_2_K_4_r=1/3}
\begin{tabular}{ccc}
\hline
\multicolumn{1}{|c|}{$s$}                    & \multicolumn{1}{c|}{DB1}           & \multicolumn{1}{c|}{DB2}           \\ \hline
\multicolumn{1}{|c|}{\multirow{3}{*}{\rotatebox[origin=c]{90}{\parbox[c]{1cm}{\centering $s=2$}}}}
		                            & \multicolumn{1}{c|}{$a_5+b_3+c_3$} & \multicolumn{1}{c|}{$a_8+b_1+c_1$}  \\ \cline{2-3}
\multicolumn{1}{|c|}{}              & \multicolumn{1}{c|}{$a_6+d_3+b_4$} & \multicolumn{1}{c|}{$a_9+d_1+b_2$}   \\ \cline{2-3}
\multicolumn{1}{|c|}{}              & \multicolumn{1}{c|}{$a_7+c_4+d_4$} & \multicolumn{1}{c|}{$a_{10}+c_2+d_2$}   \\ \hline
\multicolumn{1}{|c|}{\multirow{2}{*}{}}   & \multicolumn{1}{c|}{$b_5+c_5+d_5$} & \multicolumn{1}{c|}{$b_6+c_6+d_6$}   \\ \cline{2-3}
\multicolumn{1}{|c|}{}       & \multicolumn{1}{c|}{$a_{11}+b_6+c_6+d_6$} & \multicolumn{1}{c|}{$a_{12}+b_5+c_5+d_5$} \\ \hline
		                     &                                    &                                    \\ \cline{2-3}
\multicolumn{1}{c|}{}        &\multicolumn{1}{c|}{$Z_1=(a_1,a_2,b_1,b_2,c_1,c_2,d_1,d_2)$}                                  
                             &\multicolumn{1}{c|}{$Z_2=(a_3,a_4,b_3,b_4,c_3,c_4,d_3,d_4)$} \\ \cline{2-3}
\end{tabular}
\end{table}

\begin{table}[H]
\centering
\caption{Query table for $K=4$, $N=2$, $r_3=\frac{1}{2}$.}
\label{N_2_K_4_r=1/2}
\begin{tabular}{ccc}
\hline
\multicolumn{1}{|c|}{$s$}   & \multicolumn{1}{c|}{DB1}              & \multicolumn{1}{c|}{DB2}           \\ \hline
\multicolumn{1}{|c|}{$s=3$} & \multicolumn{1}{c|}{$a_3+b_2+c_2+d_2$} & \multicolumn{1}{c|}{$a_4+b_1+c_1+d_1$} \\ \hline
		                    &                                       &                                    \\ \cline{2-3}
\multicolumn{1}{c|}{}       &\multicolumn{1}{c|}{$Z_1=(a_1,b_1,c_1,d_1)$}&\multicolumn{1}{c|}{$Z_2=(a_2,b_2,c_2,d_2)$}\\ \cline{2-3}
\end{tabular}
\end{table}

\subsection{$K=5$, $K=10$ and $K=100$ Messages, $N=2$ Databases}

\begin{figure}[t]
	\centering
	\epsfig{file=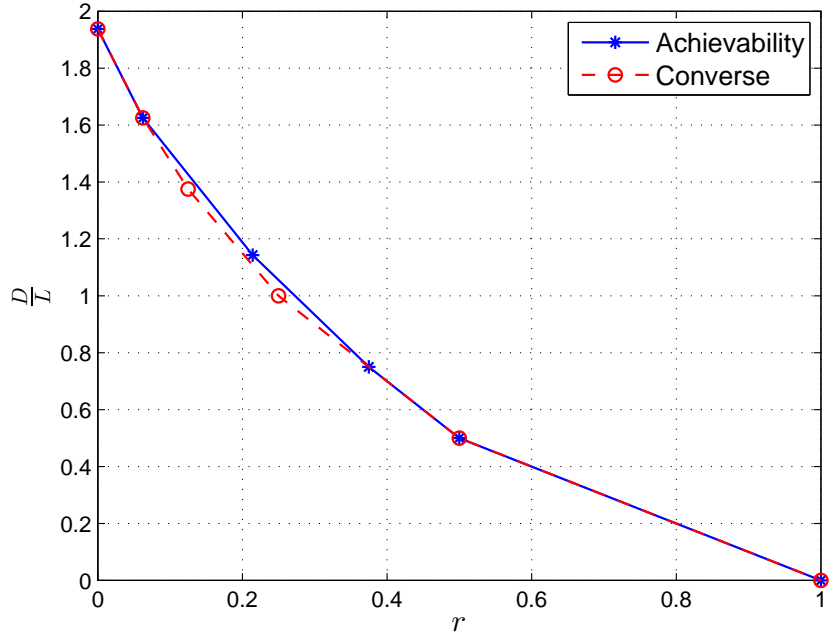,width=0.75\textwidth}
	\caption{Inner and outer bounds for $K=5$, $N=2$.}
	\label{fig_N2K5}
\end{figure}

\begin{figure}[t]
	\centering
	\epsfig{file=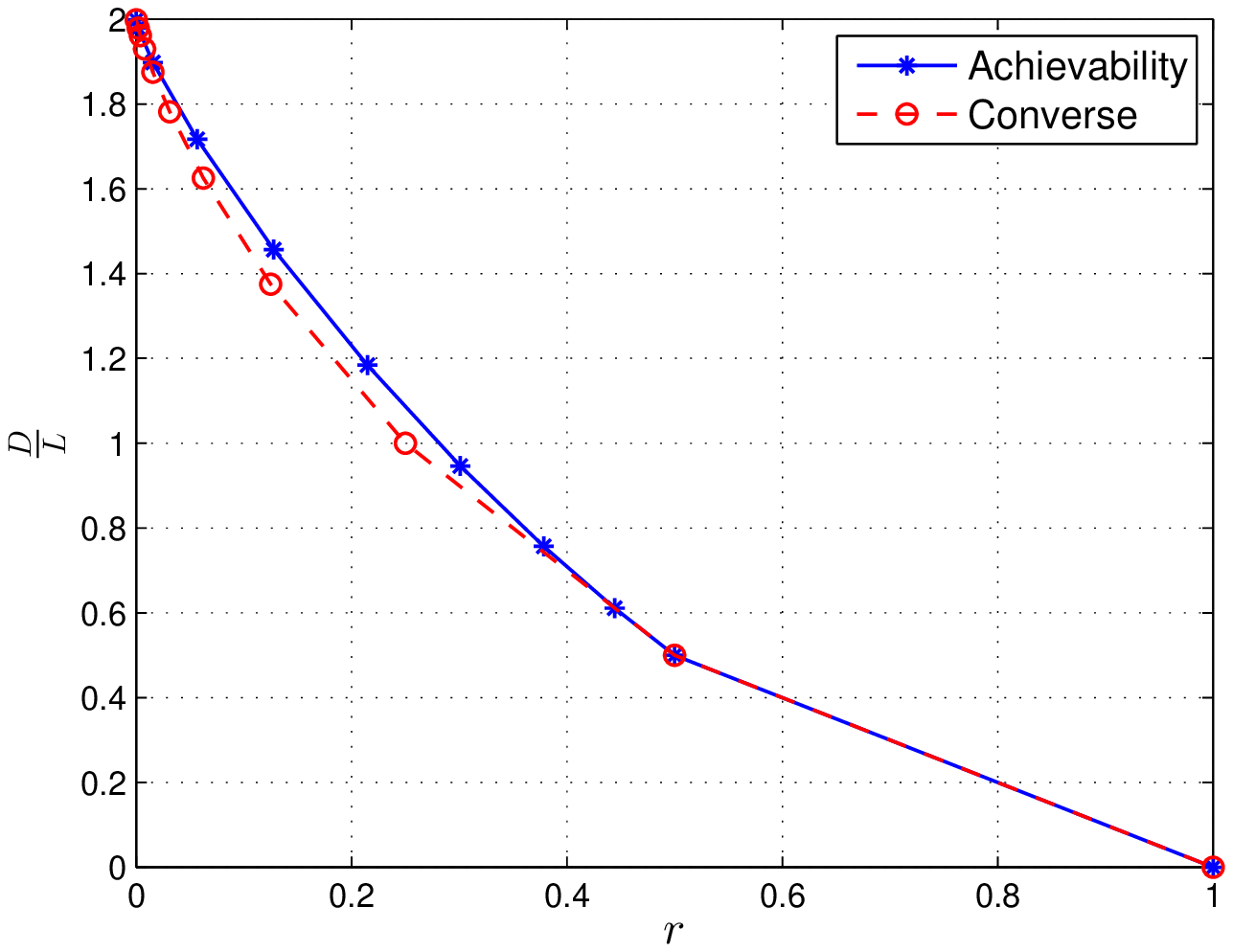,width=0.75\textwidth}
	\caption{Inner and outer bounds for $K=10$, $N=2$.}
	\label{fig_N2K10}
\end{figure}

\begin{figure}[t]
	\centering
	\epsfig{file=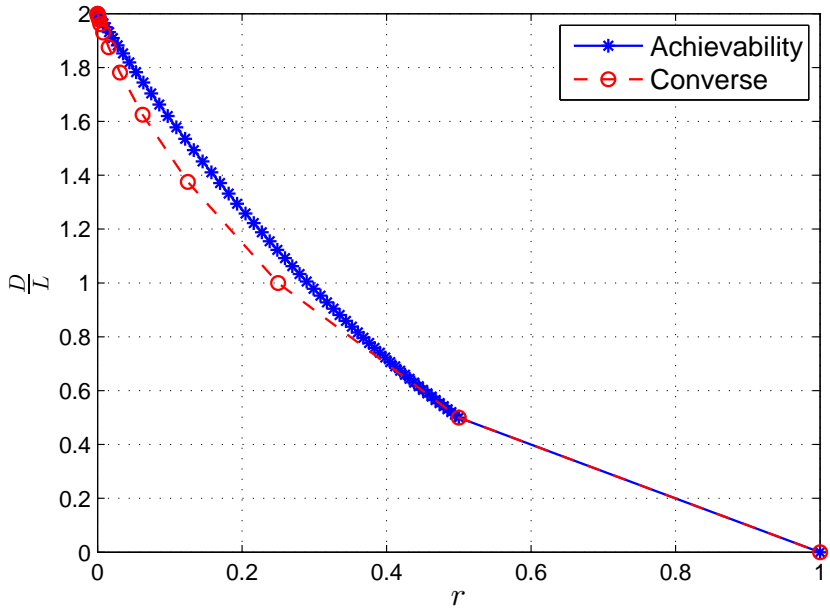,width=0.75\textwidth}
	\caption{Inner and outer bounds for $K=100$, $N=2$.}
	\label{fig_N2K100}
\end{figure}

For $N=2$, we show the numerical results for the inner and outer bounds for $K=5$, $K=10$ and $K=100$  in Figures~\ref{fig_N2K5}, \ref{fig_N2K10} and \ref{fig_N2K100}. For fixed $N$ as $K$ grows, the gap between the achievable bound and converse bound increases. This observation will be made specific in Section~\ref{Sec_gap}.

\section{Gap Analysis} \label{Sec_gap}

\begin{figure}[t]
\centering
\epsfig{file=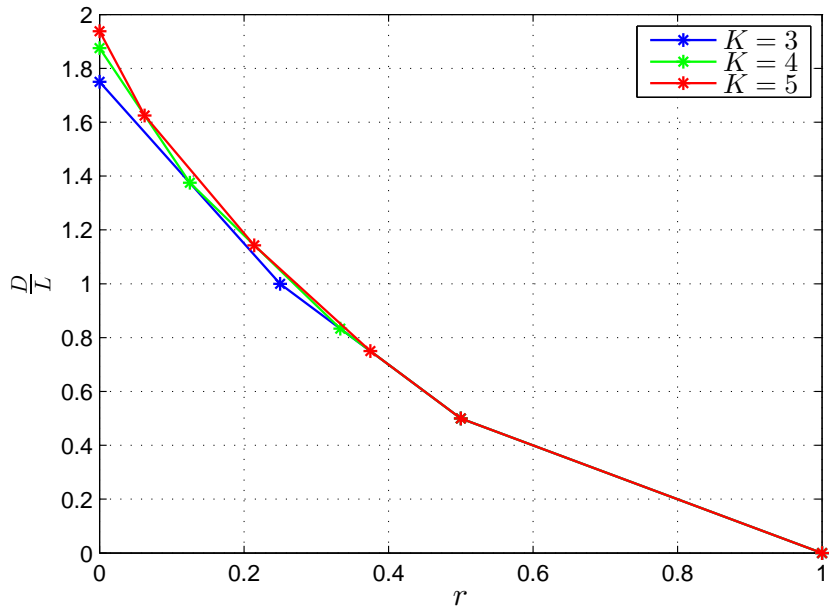,width=0.75\textwidth}
\caption{Outer bounds for $N=2$, $K=3$, $K=4$ and $K=5$.}
\label{fig_N2K345}
\end{figure}

In this section, we analyze the gap between the achievable bounds given in \eqref{eq_outer} and the converse bounds given in \eqref{eq_inner}. We first observe that for fixed number of databases $N$, as the number of messages $K$ increases, the achievable normalized download cost increases. In addition to the monotonicity, the achievable normalized download cost for $K+1$ messages has a special relationship with the achievable normalized download cost for $K$ messages. We first use an example to illustrate this property. For $N=2$, $K=3$, $K=4$, and $K=5$, the achievable bounds are shown in Figure~\ref{fig_N2K345}. The achievable bound for $K=5$ is above the achievable bound for $K=4$, and the achievable bound for $K=4$ is above the achievable bound for $K=3$. By denoting $r_s^{(K)}$ as the caching ratio with total $K$ messages and parameter $s$ (see \eqref{r_exp}), we observe that $(r^{(5)}_1, \bar{D}(r^{(5)}_1) )$ falls on the line connecting  $(r^{(4)}_0, \bar{D}(r^{(4)}_0) )$  and $(r^{(4)}_1, \bar{D}(r^{(4)}_1) )$. This observation is general, $(r_s^{(K+1)},\bar{D}(r_s^{(K+1)}) )$ falls on the line connecting $(r_{s-1}^{(K)},\bar{D}(r_{s-1}^{(K)}))$ and $(r_{s}^{(K)},\bar{D}(r_{s}^{(K)}))$. We summarize this result in the following lemma.

\begin{lemma}[Monotonicity of the achievable bounds]\label{L_monotone}
In cache-aided PIR with partially known uncoded prefetching, for fixed number of databases $N$, if the number of messages $K$ increases, then the achievable normalized download cost increases. Furthermore, we have
\begin{align}
r_s^{(K+1)} &= \alpha r_{s-1}^{(K)} + (1-\alpha) r_s^{(K)}, \label{m1} \\
\bar{D}(r_s^{(K+1)}) &=  \alpha \bar{D}(r_{s-1}^{(K)}) + (1-\alpha) \bar{D}(r_s^{(K)}),  \label{m2}
\end{align}
where $0\leq \alpha \leq 1$.
\end{lemma}
The proof of Lemma~\ref{L_monotone} is similar to \cite[Lemma~4]{wei2017fundamental}.

After showing the monotonicity of the achievable bounds, we show that as $K \rightarrow \infty$, the asymptotic upper bound for the achievable bounds is given as in the following lemma. With this asymptotic upper bound, we conclude that the worst-case gap is $\frac{5}{32}$.

\begin{lemma}[Asymptotics and the worst-case gap]\label{L_asym}
In cache-aided PIR with partially known uncoded prefetching, as $K \rightarrow \infty$, the outer bound is upper bounded by,
\begin{align}
\bar{D}(r) \leq \frac{N}{N-1}(1-r)^2
\end{align}
Hence, the worst-case gap is $\frac{5}{32}$.
\end{lemma}
The proof of Lemma~\ref{L_monotone} is provided in Section~\ref{Proof_L_asym}. 
\section{Comparisons with Other Cache-Aided PIR Models}

\begin{figure}[t]
\centering
\epsfig{file=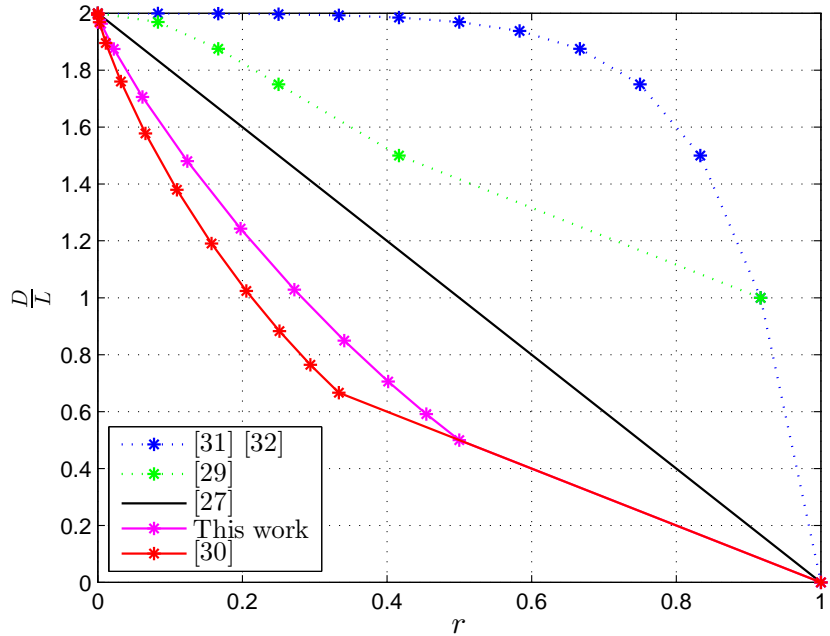,width=0.75\textwidth}
\caption{Outer bounds for $N=2$, $K=12$ for different cache-aided PIR models.}
\label{fig_cmp}
\end{figure}

In this section, we compare the normalized download costs between different cache-aided PIR models subjected to same memory size constraint. We first use an example of $N=2$ and $K=12$ (see Figure~\ref{fig_cmp}) to show the relative normalized download costs for different models. In \cite{kadhe2017private, chen2017capacity, wei2017capacity}, the user caches $M$ full messages out of total $K$ messages. In order to compare with other cache-aided PIR schemes, we use $\frac{M}{K}$ as the caching ratio. Since the PIR schemes are only reported for the corner points in \cite{kadhe2017private, chen2017capacity, wei2017capacity}, we use dotted lines to connect the corner points. For \cite{wei2017fundamental,tandon2017capacity} and this work, since we can apply memory-sharing to achieve the download costs between the corner points, we use solid lines to connect the corner points.

We first compare references \cite{kadhe2017private, chen2017capacity, wei2017capacity}, in which the user caches $M$ full messages out of $K$ messages and the databases are (partially) unaware. In \cite{chen2017capacity, wei2017capacity}, the user not only wishes to protect the privacy of the desired messages but also wishes to protect the privacy of the cached messages. Note that the other works (\cite{kadhe2017private, wei2017fundamental,tandon2017capacity} and this work) only consider to protect the privacy of the desired messages. Since the message privacy constraint is less restricted, reference \cite{kadhe2017private} achieves lower normalized download cost than references \cite{chen2017capacity, wei2017capacity}. The main difference between \cite{wei2017capacity} and \cite{chen2017capacity} is that the databases are totally unaware of the cached $M$ messages as in \cite{chen2017capacity} or the $n$th database is aware of some of $M$ messages cached from the $n$th database as in \cite{wei2017capacity}. Interestingly, these two models result in the same normalized download costs. Although the $n$th database's awareness of some cached messages might increase the download cost, at the same time the user does not need to protect the privacy of these known messages from the $n$th database, which might reduce the download cost.

We then compare references \cite{tandon2017capacity,wei2017fundamental} and this work. The main difference between these three works is the different level of awareness of the side information the user cached. Reference \cite{tandon2017capacity} considers that all the databases are aware of the side information the user cached. In contrast, reference \cite{wei2017fundamental} considers that all the databases are unaware of the side information. This work considers that ths $n$th database is aware of the side information cached from the $n$th database. Reference \cite[Corollary~1]{wei2017fundamental} shows the unawareness gain. Therefore, reference \cite{wei2017fundamental} achieves lower normalized download cost than \cite{tandon2017capacity}. The same proof technique in \cite[Corollary~1]{wei2017fundamental} can also show the partially unawareness gain. Therefore, this work also achieves lower normalized download cost than \cite{tandon2017capacity}. Since these three works consider only the privacy of the desired message, different from \cite{chen2017capacity, wei2017capacity}, reference \cite{wei2017fundamental} achieves lower normalized download cost than this work.

\section{Conclusion}
In this paper, we studied the cache-aided PIR problem from $N$ non-communicating and replicated databases, when the cache stores uncoded bits that are partially known to the databases. We determined inner and outer bounds for the optimal normalized download cost $D^*(r)$ as a function of the total number of messages $K$, the number of databases $N$, and the caching ratio $r$. Both inner and outer bounds are piece-wise linear functions in $r$ (for fixed $N$, $K$) that consist of $K$ line segments. The bounds match in two specific regimes: the very low caching ratio regime, i.e., $r \leq \frac{1}{N^{K-1}}$, and the very high caching ratio regime, where $r \geq \frac{K-2}{N^2-3N+KN}$. As a direct corollary for this result, we characterized the exact tradeoff between the download cost and the caching ratio for $K=3$. For general $K$, $N$, and $r$, we showed that the largest gap between the achievability and the converse bounds is $\frac{5}{32}$. The achievable scheme extends the greedy scheme in \cite{JafarPIR} so that it starts with exploiting the cache bits as side information. For fixed $K$, $N$, there are $K-1$ non-degenerate corner points. These points differ in the number of cached bits that contribute in generating one side information equation. The achievability for the remaining caching ratios is done by memory-sharing between the two adjacent corner points that enclose that caching ratio $r$. For the converse, we extend the induction-based techniques in \cite{JafarPIR,wei2017fundamental} to account for the availability of uncoded and partially prefetched side information at the retriever. The converse proof hinges on developing $K$ lower bounds on the length of the undesired portion of the answer string. By applying induction on each bound separately, we obtain the piece-wise linear inner bound.
\section{Appendix}

\subsection{Proof of Lemma~\ref{L_asym}}\label{Proof_L_asym}
\begin{Proof}
From \eqref{eq_outer}, we rewrite $\bar{D}(r_s)$ as
\begin{align}
\bar{D}(r_s)&= \frac{\sum_{i=0}^{K-1-s} \binom{K}{s+1+i}(N-1)^{i+1} }{\binom{K-2}{s-1}+\sum_{i=0}^{K-1-s} \binom{K-1}{s+i}(N-1)^{i+1}}\\
&=\frac{\frac{\sum_{i=0}^{K-1-s} \binom{K}{s+1+i}(N-1)^{i+1}}{\sum_{i=0}^{K-1-s} \binom{K-1}{s+i}(N-1)^{i+1}}}{\frac{\binom{K-2}{s-1}}{\sum_{i=0}^{K-1-s} \binom{K-1}{s+i}(N-1)^{i+1}}+1} =\frac{\psi_1(N,K,s)}{\psi_2(N,K,s)+1}. \label{ub_gap}
\end{align}

Let $\lambda=\frac{s}{K}$. We first upper bound $\psi_1(N,K,s)$,
\begin{align}
\psi_1(N,K,s)&=\frac{\sum_{i=0}^{K-1-s} \binom{K}{s+1+i}(N-1)^{i+1}}{\sum_{i=0}^{K-1-s} \binom{K-1}{s+i}(N-1)^{i+1}} \\
&=\frac{\sum_{i=0}^{K-1-s} \frac{K}{s+1+i}\binom{K-1}{s+i}(N-1)^{i+1}}{\sum_{i=0}^{K-1-s}
	\binom{K-1}{s+i}(N-1)^{i+1}}\\
& \leq \frac{\sum_{i=0}^{K-1-s} \frac{K}{s}\binom{K-1}{s+i}(N-1)^{i+1}}{\sum_{i=0}^{K-1-s}
	\binom{K-1}{s+i}(N-1)^{i+1}}=\frac{1}{\lambda}. \label{nn3}
\end{align}

We then upper bound the reciprocal of $\psi_2(N,K,s)$ as,
\begin{align}
\frac{1}{\psi_2(N,K,s)}&=\sum_{i=0}^{K-1-s} \frac{\binom{K-1}{s+i}(N-1)^{i+1}}{\binom{K-2}{s-1}}\\
&\label{gap_equality}=(N-1)\sum_{i=0}^{K-1-s} \frac{(K-1)(K-1-s)(K-2-s) \cdots (K-i-s)}{s(s+1)(s+2)\cdots (s+i)} (N-1)^i \\
&\leq (N-1)\sum_{i=0}^{K-1-s} \frac{K(K-s)^{i}}{s^{i+1}}(N-1)^i\\
&=\frac{(N-1)}{\lambda} \sum_{i=0}^{(1-\lambda)K-1} \left(\frac{(1-\lambda)(N-1)}{\lambda}\right)^i.
\end{align}
When $\lambda > 1-\frac{1}{N}$, $\frac{(1-\lambda)(N-1)}{\lambda} <1$. As $K \rightarrow \infty$, $\frac{1}{\psi_2(N,K,s)}$ is upper bounded by
\begin{align} \label{nn2}
\lim_{K \rightarrow \infty} \frac{1}{\psi_2(N,K,s)} &\leq\frac{N-1}{\lambda} \sum_{i=0}^{\infty} \left(\frac{(1-\lambda)(N-1)}{\lambda}\right)^i \\
&=\frac{N-1}{\lambda}\cdot\frac{1}{1-\frac{(1-\lambda)(N-1)}{\lambda}} =\frac{N-1}{N\lambda-(N-1)}.
\end{align}
	
Now, we lower bound \eqref{gap_equality} by keeping the first $\epsilon K$ terms in the sum for any $\epsilon$ such that $0 < \epsilon <1-\lambda$,
\begin{align}
\frac{1}{\psi_2(N,K,s)} &\geq (N-1)\sum_{i=0}^{\epsilon K} \frac{(K-1)(K-1-s)(K-2-s) \cdots (K-i-s)}{s(s+1)(s+2)\cdots (s+i)}(N-1)^i \\
&\geq (N-1)\sum_{i=0}^{\epsilon K} \frac{(K-1)(K-\epsilon K-s)^{i}}{(s+\epsilon K)^{i+1}} (N-1)^i\\
&= (N-1)\sum_{i=0}^{\epsilon K} \frac{(1-\frac{1}{K})((1-(\lambda+\epsilon))^{i}}{(\lambda+\epsilon)^{i+1}}(N-1)^i.
\end{align}
As $K \rightarrow \infty$, for any $0 < \epsilon <1-\lambda$, we have
\begin{align} \label{nn1}
\lim_{K \rightarrow \infty} \frac{1}{\psi_2(N,K,s)} &\geq\frac{N-1}{\lambda+\epsilon} \sum_{i=0}^{\infty} \left(\frac{(1-(\lambda+\epsilon))(N-1)}{\lambda+\epsilon}\right)^i \\
&=\frac{N-1}{N(\lambda+\epsilon)-(N-1)}.
\end{align}
From \eqref{nn1} and \eqref{nn2}, as $K \rightarrow \infty$, by picking $\epsilon \rightarrow 0$, we have
\begin{align} \label{nn4}
\psi_2(N,K,s) \rightarrow \frac{N}{N-1}\lambda-1.
\end{align}
	
Furthermore, as $K \rightarrow \infty$, $r_s$ converges to
\begin{align}
r_s \rightarrow r&= \lim_{K \rightarrow \infty}\frac{\binom{K-2}{s-1} }{\binom{K-2}{s-1}+\sum_{i=0}^{K-1-s} \binom{K-1}{s+i}(N-1)^{i+1}} \\
&=\lim_{K \rightarrow \infty} \frac{\psi_2(N,K,s)}{\psi_2(N,K,s)+1}\\
&=\frac{N\lambda-(N-1)}{N\lambda}=1-\left(1-\frac{1}{N}\right) \frac{1}{\lambda}.
\end{align}
Note that if $\lambda=1-\frac{1}{N}$, then $r=0$, while if $\lambda=1$, then $r=\frac{1}{N}$. Since we now consider the gap in the region of $0 \leq r \leq \frac{1}{N}$, without loss of generality, we consider $\lambda>1-\frac{1}{N}$. We express $\lambda$ as
\begin{align} \label{nn5}
\lambda=\frac{1-\frac{1}{N}}{1-r}.
\end{align}

Continuing \eqref{ub_gap}, by using \eqref{nn3}, \eqref{nn4} and \eqref{nn5}, we have the following upper bound on $\bar{D}(r)$
\begin{align}
\bar{D}(r) \leq \frac{\frac{1}{\lambda}}{\frac{N}{N-1}\lambda} =\frac{1}{\lambda^2}\left(1-\frac{1}{N}\right)=\frac{N}{N-1} (1-r)^2.  \label{nn6}
\end{align}
	
Now, we compare the inner bound in \eqref{eq_inner} with the outer bound derived in \eqref{nn6}. Note that the inner bound in \eqref{eq_inner} consists of $K$ line segments, and these $K$ line segments intersect at the following $K-1$ points given by,
\begin{align} \label{nn7}
\tilde{r}_i=\frac{1}{N^i}, \quad i=1,\cdots,K-1.
\end{align}
As $i$ increases, $\tilde{r}_i$ concentrates to $r=0$. Therefore, for these $K$ line segments, we only need to consider small number of them for the worst-gap analysis. Denote the gap between the inner and the outer bounds by $\Delta(N,K,r)$. We note that the gap $\Delta(N,\infty,r)$ is a piece-wise convex function for $0 \leq r \leq 1$ since it is the difference between a convex function $\bar{D}(r)$ and a piece-wise linear function. Hence, the maximizing caching ratio for the gap exists exactly at the corner points $\tilde{r}_i$ and it suffices to examine the gap at these corner points.
	
For the outer bound, by plugging \eqref{nn7} into \eqref{nn6}, we have
\begin{align}
\bar{D}(\tilde{r}_i)\leq \frac{N}{N-1}\left(1-\frac{1}{N^i}\right)^2 =\frac{1-(\frac{1}{N})^i}{1-\frac{1}{N}}\left(1-\frac{1}{N^i}\right).
\end{align}

Furthermore, for the inner bound, we have
\begin{align}
\tilde{D}(\tilde{r}_i)=&(1-r_i)\left(1+\frac{1}{N}+\cdots+\frac{1}{N^{i}}\right)-r_i\left(1-\frac{1}{N}\right)\left(i+\frac{(i-1)}{N}+\cdots+\frac{1}{N^{i-1}}\right)\\
=&-r_i\left[\left(1+\frac{1}{N}+\cdots+\frac{1}{N^{i}}\right)+\left(1-\frac{1}{N}\right)\left(i+\frac{(i-1)}{N}+\cdots+\frac{1}{N^{i-1}}\right)\right]\notag\\
&+\left(1+\frac{1}{N}+\cdots+\frac{1}{N^{i}}\right)\\
=&-r_i(i+1)+\left(1+\frac{1}{N}+\cdots+\frac{1}{N^{i}}\right)\\
=&\frac{1-(\frac{1}{N})^{i+1}}{1-\frac{1}{N}}-r_i(i+1) =\frac{1-(\frac{1}{N})^{i+1}}{1-\frac{1}{N}}-\frac{i+1}{N^i}
\end{align}

Consequently, we can upper bound the asymptotic gap at the corner point $\tilde{r}_i$ as
\begin{align}
\Delta(N,\infty,\tilde{r}_i)= \bar{D}(\tilde{r}_i)-\tilde{D}(\tilde{r}_i)\leq \frac{1}{N^i}\left[i-\frac{1-(\frac{1}{N})^i}{1-\frac{1}{N}}\right]
\end{align}
	Hence, $\Delta(N,\infty,\tilde{r}_i)$ is monotonically decreasing in $N$. Therefore,
	\begin{align}
	\Delta(N,K,r) \leq \Delta(2,\infty,r) \leq \: \max_i \:\: \frac{1}{2^i}\left[i-\frac{1-(\frac{1}{2})^i}{1-\frac{1}{2}}\right]
	\end{align}
	For the case $N=2$, we note that all the inner bounds after the $7$th corner point are concentrated around $r=0$ since $\tilde{r}_i \leq \frac{1}{128}$ for $i \geq 7$. Therefore, it suffices to characterize the gap only for the first $7$ corner points. Considering the $7$th corner point which corresponds to $\tilde{r}_6=\frac{1}{128}$, and $\bar{D}(r) \leq 2$ trivially for all $r$, and $\tilde{D}(\frac{1}{128})=1.9297$. Hence, $\Delta(2,\infty,r) \leq 0.07$, for $r \leq \frac{1}{127}$. Now, we focus on calculating the gap at $\tilde{r}_i$, $i=1, \cdots, 7$. Examining all the corner points, we see that $r=\frac{1}{8}$ is the maximizing caching ratio for the gap (corresponding to $i=3$), and $\Delta(2,\infty,\frac{1}{8}) \leq \frac{5}{32}$, which is the worst-case gap.
\end{Proof}

\bibliographystyle{unsrt}
\bibliography{references}

\begin{thebibliography}{10}

\bibitem{ChorPIR}
B.~Chor, E.~Kushilevitz, O.~Goldreich, and M.~Sudan.
\newblock Private information retrieval.
\newblock {\em Journal of the ACM}, 45(6):965--981, 1998.

\bibitem{PIRsurvey2004}
W.~Gasarch.
\newblock A survey on private information retrieval.
\newblock In {\em Bulletin of the EATCS}, 2004.

\bibitem{cachin1999computationally}
C.~Cachin, S.~Micali, and M.~Stadler.
\newblock Computationally private information retrieval with polylogarithmic
  communication.
\newblock In {\em International Conference on the Theory and Applications of
  Cryptographic Techniques}. Springer, 1999.

\bibitem{ostrovsky2007survey}
R.~Ostrovsky and W.~Skeith III.
\newblock A survey of single-database private information retrieval: Techniques
  and applications.
\newblock In {\em International Workshop on Public Key Cryptography}, pages
  393--411. Springer, 2007.

\bibitem{yekhanin2010private}
S.~Yekhanin.
\newblock Private information retrieval.
\newblock {\em Communications of the ACM}, 53(4):68--73, 2010.

\bibitem{RamchandranPIR}
N.~B. Shah, K.~V. Rashmi, and K.~Ramchandran.
\newblock One extra bit of download ensures perfectly private information
  retrieval.
\newblock In {\em IEEE ISIT}, June 2014.

\bibitem{unsynchonizedPIR}
G.~Fanti and K.~Ramchandran.
\newblock Efficient private information retrieval over unsynchronized
  databases.
\newblock {\em IEEE Journal of Selected Topics in Signal Processing},
  9(7):1229--1239, October 2015.

\bibitem{YamamotoPIR}
T.~Chan, S.~Ho, and H.~Yamamoto.
\newblock Private information retrieval for coded storage.
\newblock In {\em IEEE ISIT}, June 2015.

\bibitem{VardyConf2015}
A.~Fazeli, A.~Vardy, and E.~Yaakobi.
\newblock Codes for distributed {PIR} with low storage overhead.
\newblock In {\em IEEE ISIT}, June 2015.

\bibitem{RazanPIR}
R.~Tajeddine and S.~El Rouayheb.
\newblock Private information retrieval from {MDS} coded data in distributed
  storage systems.
\newblock In {\em IEEE ISIT}, July 2016.

\bibitem{JafarConf2016}
H.~Sun and S.~A. Jafar.
\newblock The capacity of private information retrieval.
\newblock In {\em IEEE Globecom}, December 2016.

\bibitem{JafarPIR}
H.~Sun and S.~A. Jafar.
\newblock The capacity of private information retrieval.
\newblock {\em IEEE Transactions on Information Theory}, 63(7):4075--4088, July
  2017.

\bibitem{JafarColluding}
H.~Sun and S.~Jafar.
\newblock The capacity of robust private information retrieval with colluding
  databases.
\newblock 2016.
\newblock Available at arXiv:1605.00635.

\bibitem{symmetricPIR}
H.~Sun and S.~Jafar.
\newblock The capacity of symmetric private information retrieval.
\newblock 2016.
\newblock Available at arXiv:1606.08828.

\bibitem{KarimCoded}
K.~Banawan and S.~Ulukus.
\newblock The capacity of private information retrieval from coded databases.
\newblock {\em IEEE Transactions on Information Theory}.
\newblock Submitted September 2016. Also available at arXiv:1609.08138.

\bibitem{arbmsgPIR}
H.~Sun and S.~Jafar.
\newblock Optimal download cost of private information retrieval for arbitrary
  message length.
\newblock 2016.
\newblock Available at arXiv:1610.03048.

\bibitem{codedsymmetric}
Q.~Wang and M.~Skoglund.
\newblock Symmetric private information retrieval for {MDS} coded distributed
  storage.
\newblock 2016.
\newblock Available at arXiv:1610.04530.

\bibitem{MultiroundPIR}
H.~Sun and S.~Jafar.
\newblock Multiround private information retrieval: Capacity and storage
  overhead.
\newblock 2016.
\newblock Available at arXiv:1611.02257.

\bibitem{codedcolluded}
R.~Freij-Hollanti, O.~Gnilke, C.~Hollanti, and D.~Karpuk.
\newblock Private information retrieval from coded databases with colluding
  servers.
\newblock 2016.
\newblock Available at arXiv:1611.02062.

\bibitem{codedcolludedJafar}
H.~Sun and S.~Jafar.
\newblock Private information retrieval from {MDS} coded data with colluding
  servers: Settling a conjecture by {F}reij-{H}ollanti et al.
\newblock 2017.
\newblock Available at arXiv: 1701.07807.

\bibitem{arbitraryCollusion}
R.~Tajeddine, O.~W. Gnilke, D.~Karpuk, R.~Freij-Hollanti, C.~Hollanti, and
  S.~El Rouayheb.
\newblock Private information retrieval schemes for coded data with arbitrary
  collusion patterns.
\newblock 2017.
\newblock Available at arXiv:1701.07636.

\bibitem{MPIRjournal}
K.~Banawan and S.~Ulukus.
\newblock Multi-message private information retrieval: Capacity results and
  near-optimal schemes.
\newblock {\em IEEE Transactions on Information Theory}.
\newblock Submitted February 2017. Also available at arXiv:1702.01739.

\bibitem{codedcolludingZhang}
Y.~Zhang and G.~Ge.
\newblock A general private information retrieval scheme for {MDS} coded
  databases with colluding servers.
\newblock 2017.
\newblock Available at arXiv: 1704.06785.

\bibitem{MPIRcodedcolludingZhang}
Y.~Zhang and G.~Ge.
\newblock Multi-file private information retrieval from {MDS} coded databases
  with colluding servers.
\newblock 2017.
\newblock Available at arXiv: 1705.03186.

\bibitem{BPIRjournal}
K.~Banawan and S.~Ulukus.
\newblock The capacity of private information retrieval from {B}yzantine and
  colluding databases.
\newblock {\em IEEE Transactions on Information Theory}.
\newblock Submitted June 2017. Also available at arXiv:1706.01442.

\bibitem{symmetricByzantine}
Q.~Wang and M.~Skoglund.
\newblock Secure symmetric private information retrieval from colluding
  databases with adversaries.
\newblock 2017.
\newblock Available at arXiv:1707.02152.

\bibitem{tandon2017capacity}
R.~Tandon.
\newblock The capacity of cache aided private information retrieval.
\newblock 2017.
\newblock Available at arXiv: 1706.07035.

\bibitem{wang2017linear}
Q.~Wang and M.~Skoglund.
\newblock Linear symmetric private information retrieval for {MDS} coded
  distributed storage with colluding servers.
\newblock 2017.
\newblock Available at arXiv:1708.05673.

\bibitem{kadhe2017private}
S.~Kadhe, B.~Garcia, A.~Heidarzadeh, S.~El Rouayheb, and A.~Sprintson.
\newblock Private information retrieval with side information.
\newblock 2017.
\newblock Available at arXiv:1709.00112.

\bibitem{wei2017fundamental}
Y.-P. Wei, K.~Banawan, and S.~Ulukus.
\newblock Fundamental limits of cache-aided private information retrieval with
  unknown and uncoded prefetching.
\newblock 2017.
\newblock Available at arXiv:1709.01056.

\bibitem{chen2017capacity}
Z.~Chen, Z.~Wang, and S.~Jafar.
\newblock The capacity of private information retrieval with private side
  information.
\newblock 2017.
\newblock Available at arXiv:1709.03022.

\bibitem{wei2017capacity}
Y.-P. Wei, K.~Banawan, and S.~Ulukus.
\newblock The capacity of private information retrieval with partially known
  private side information.
\newblock 2017.
\newblock Available at arXiv:1710.00809.

\bibitem{sun2017_computation}
H.~Sun and S.~A. Jafar.
\newblock The capacity of private computation.
\newblock 2017.
\newblock Available at arXiv:1710.11098.

\bibitem{mirmohseni2017private}
M.~Mirmohseni and M.~A. Maddah-Ali.
\newblock Private function retrieval.
\newblock 2017.
\newblock Available at arXiv:1711.04677.

\bibitem{abdul2017private}
M.~Abdul-Wahid, F.~Almoualem, D.~Kumar, and R.~Tandon.
\newblock Private information retrieval from storage constrained
  databases--coded caching meets {PIR}.
\newblock 2017.
\newblock Available at arXiv:1711.05244.

\bibitem{maddah2014fundamental}
M.~A. Maddah-Ali and U.~Niesen.
\newblock Fundamental limits of caching.
\newblock {\em IEEE Transactions on Information Theory}, 60(5):2856--2867, May
  2014.

\bibitem{maddah2015decentralized}
M.~A. Maddah-Ali and U.~Niesen.
\newblock Decentralized coded caching attains order-optimal memory-rate
  tradeoff.
\newblock {\em IEEE/ACM Transactions on Networking}, 23(4):1029--1040, August
  2015.

\bibitem{pedarsani2016online}
R.~Pedarsani, M.~A. Maddah-Ali, and U.~Niesen.
\newblock Online coded caching.
\newblock {\em IEEE/ACM Transactions on Networking}, 24(2):836--845, April
  2016.

\bibitem{sengupta2015fundamental}
A.~Sengupta, R.~Tandon, and T.~C. Clancy.
\newblock Fundamental limits of caching with secure delivery.
\newblock {\em IEEE Transactions on Information Forensics and Security},
  10(2):355--370, February 2015.

\bibitem{ji2016fundamental}
M.~Ji, G.~Caire, and A.~F. Molisch.
\newblock Fundamental limits of caching in wireless {D2D} networks.
\newblock {\em IEEE Transactions on Information Theory}, 62(2):849--869,
  February 2016.

\bibitem{ghasemi2017improved}
H.~Ghasemi and A.~Ramamoorthy.
\newblock Improved lower bounds for coded caching.
\newblock {\em IEEE Transactions on Information Theory}, 63(7):4388--4413, July
  2017.

\bibitem{ji2017order}
M.~Ji, A.~M. Tulino, J.~Llorca, and G.~Caire.
\newblock Order-optimal rate of caching and coded multicasting with random
  demands.
\newblock {\em IEEE Transactions on Information Theory}, 63(6):3923--3949, June
  2017.

\bibitem{timo2015joint}
R.~Timo and M.~Wigger.
\newblock Joint cache-channel coding over erasure broadcast channels.
\newblock In {\em IEEE ISWCS}, August 2015.

\bibitem{shanmugam2016finite}
K.~Shanmugam, M.~Ji, A.~M. Tulino, J.~Llorca, and A.~G. Dimakis.
\newblock Finite-length analysis of caching-aided coded multicasting.
\newblock {\em IEEE Transactions on Information Theory}, 62(10):5524--5537,
  October 2016.

\bibitem{zhang2017fundamental}
J.~Zhang and P.~Elia.
\newblock Fundamental limits of cache-aided wireless {BC}: {I}nterplay of
  coded-caching and {CSIT} feedback.
\newblock {\em IEEE Transactions on Information Theory}, 63(5):3142--3160, May
  2017.

\bibitem{gregori2016wireless}
M.~Gregori, J.~G{\'o}mez-Vilardeb{\'o}, J.~Matamoros, and D.~G{\"u}nd{\"u}z.
\newblock Wireless content caching for small cell and {D2D} networks.
\newblock {\em IEEE Journal on Selected Areas in Communications},
  34(5):1222--1234, May 2016.

\bibitem{amiri2017fundamental}
M.~M. Amiri and D.~G{\"u}nd{\"u}z.
\newblock Fundamental limits of coded caching: Improved delivery rate-cache
  capacity tradeoff.
\newblock {\em IEEE Transactions on Communications}, 65(2):806--815, February
  2017.

\bibitem{tian2016caching}
C.~Tian and J.~Chen.
\newblock Caching and delivery via interference elimination.
\newblock In {\em IEEE ISIT}, July 2016.

\bibitem{wan2016optimality}
K.~Wan, D.~Tuninetti, and P.~Piantanida.
\newblock On the optimality of uncoded cache placement.
\newblock In {\em IEEE ITW}, September 2016.

\bibitem{yu2016exact}
Q.~Yu, M.~A. Maddah-Ali, and A.~S. Avestimehr.
\newblock The exact rate-memory tradeoff for caching with uncoded prefetching.
\newblock In {\em IEEE ISIT}, June 2017.

\bibitem{zewail2016fundamental}
A.~A. Zewail and A.~Yener.
\newblock Fundamental limits of secure device-to-device coded caching.
\newblock In {\em IEEE Asilomar Conference on Signals, Systems, and Computers},
  October 2016.

\bibitem{naderializadeh2017optimality}
N.~Naderializadeh, M.~A. Maddah-Ali, and A.~S. Avestimehr.
\newblock On the optimality of separation between caching and delivery in
  general cache networks.
\newblock 2017.
\newblock Available at arXiv:1701.05881.

\bibitem{bidokhti2017benefits}
S.~S. Bidokhti, M.~Wigger, and A.~Yener.
\newblock Benefits of cache assignment on degraded broadcast channels.
\newblock 2017.
\newblock Available at arXiv:1702.08044.

\bibitem{tang2017low}
L.~Tang and A.~Ramamoorthy.
\newblock Low subpacketization schemes for coded caching.
\newblock 2017.
\newblock Available at arXiv:1706.00101.

\bibitem{xiang2017cache}
L.~Xiang, D.~W.~K. Ng, R.~Schober, and V.~W.~S. Wong.
\newblock Cache-enabled physical layer security for video streaming in
  backhaul-limited cellular networks.
\newblock {\em IEEE Transactions on Wireless Communications}, November 2017.

\end{thebibliography}
\end{document}